\documentclass[twocolumn]{article}

\usepackage{amsmath,amssymb,amsfonts,amsthm,enumerate,multirow,mathtools}
\usepackage{color}
\usepackage{siunitx}
\usepackage{parskip}
\usepackage[hidelinks]{hyperref}
\usepackage{placeins} 
\usepackage{booktabs}
\usepackage{algorithm,algpseudocode}
\usepackage[affil-it]{authblk}
\usepackage[margin=2.5cm]{geometry}
\usepackage[numbers,sort&compress]{natbib}
\usepackage{xfrac}
\usepackage{subcaption}

\newcommand{\pmat}[1]{\begin{bmatrix}#1\end{bmatrix}}

\newcommand{\Transp}{\mathsf{T}}
\DeclareMathOperator*{\argmax}{argmax}
\newcommand{\red}[1]{{\color{red}#1}}
\newcommand{\blue}[1]{{\color{blue}#1}}
\newcommand{\releps}{\langle\epsilon\rangle}

\newcolumntype{R}[1]{>{\raggedleft\let\newline\\\arraybackslash\hspace{0pt}}m{#1}}

\newcolumntype{L}[1]{>{\raggedright\let\newline\\\arraybackslash\hspace{0pt}}m{#1}}
\newcolumntype{C}[1]{>{\centering\let\newline\\\arraybackslash\hspace{0pt}}m{#1}}

\title{Bayesian Non-parametric Bragg-edge Fitting for Neutron Transmission Strain Imaging}

\author[1]{Johannes Hendriks}
\author[1]{Nicholas O'Dell}
\author[1]{Adrian Wills}
\author[2]{Anton Tremsin}
\author[1]{Christopher Wensrich}
\author[3]{Takenao Shinohara}

\affil[1]{School of Engineering, The University of Newcastle, Callaghan NSW 2308, Australia}
\affil[2]{Space Sciences Laboratory, University of California, Berkeley CA 94720, USA}
\affil[3]{Materials and Life Sciences Facility, Japan Proton Accelerator Research Complex, Tokai-mura, Ibaraki, Japan}

\begin{document}

\twocolumn[
\maketitle
\begin{@twocolumnfalse}
\begin{abstract}
Energy resolved neutron transmission techniques can provide high-resolution images of strain within polycrystalline samples allowing the study of residual strain and stress in engineered components. Strain is estimated from such data by analysing features known as Bragg-edges for which several methods exist. It is important for these methods to provide both accurate estimates of strain and an accurate quantification the associated uncertainty. Our contribution is twofold. First, we present a numerical simulation analysis of these existing methods, which shows that the most accurate estimates of strain are provided by a method that provides inaccurate estimates of certainty. Second, a novel Bayesian non-parametric method for estimating strain from Bragg-edges is presented. The numerical simulation analysis indicates that this method provides both competitive estimates of strain and accurate quantification of certainty, two demonstrations on experimental data are then presented.

\vspace{5mm}
\end{abstract}
\end{@twocolumnfalse}]



\section{Introduction and Motivation} 
\label{sec:introduction}
Energy resolved neutron transmission techniques can provide high-resolution images of strain within polycrystalline samples \citep{santisteban02,tremsin12} by analysing features known as Bragg-edges.
These strain images can be used to study the residual strain, and hence stress, within engineering components.
Residual stresses are those that remain after the applied load is removed, for example, due to heat treatment, or plastic deformation \citep{withers2001bresidual}.
The presence of residual stress can have a significant and unintended impact on a component's mechanical performance --- in particular its fatigue life.

Modern microchannel plate detectors (MCP) \citep{tremsin11}, in combination with a spallation neutron source, can measure the transmission of neutrons as a function of wavelength over a 512 by 512 grid of $\SI{55}{\micro\meter}$ pixels. Strain is estimated from these measurements by analysing the relative shift in the location of Bragg-edges.
Bragg-edges are a sudden increase in the transmission as a function of wavelength. This sudden increase occurs when the diffraction angle $2\theta$ reaches $180^\circ$ beyond which no further coherent scattering can occur. The location of these edges, $\lambda_{hkl}$, is given by Bragg's law \citep{bragg1913reflection}, $\lambda_{hkl}=2d_{hkl}\sin\theta$, and is related to strain through
\begin{equation}\label{eq:relative_strain}
    \langle \epsilon \rangle = \frac{d_{hkl} - d_0}{d_0},
\end{equation}
where $d_{hkl}$ is the lattice spacing and $d_0$ is the equivalent lattice spacing in a stress-free sample.

Several methods exist for estimating strain by analysing a single Bragg-edge in the measured transmission --- presented by \citet{santisteban2001time}, \citet{tremsin2016investigation}, and \citet{ramadhan2018neutron}. These methods are reviewed in Section~\ref{sec:existing_approaches}. 
Given a measured transmission these methods produce:
\begin{enumerate}
    \item An estimate of the strain,
    \item A quantification of their certainty in this estimate; this could be predicted confidence limits or standard deviation.
\end{enumerate}
The measurements of the transmission are corrupted by noise. It is commonplace to average over several pixels either using a running average or by grouping pixels together into macro pixels. However, this averaging results in unwanted smoothing and should be kept to a minimum. Therefore, in order to confidently use these methods to study residual stress and make engineering decisions, it is important that they are accurate even when the data is noisy. 
Further, it is important that the quantification of certainty provided by the methods is reliable, allowing the user to make appropriate decisions about the validity of the estimated strains.
Additionally, if the methods are capable of providing accurate estimates of strain even from noisier data, then less time is required for data acquisition.

The ability of these methods to produce accurate estimates of strain along with reliable confidence estimates from noisy data is particularly important for strain tomography and for dynamic measurement strain. Dynamic measurements of strain during in-situ material loading have measurement acquisition times limited by the process itself and these short acquisition times lead to noisy data \cite{mostafavi2017dynamic}.
Strain tomography is analogous to computed medical tomography and reconstructs the full triaxial strain field within a sample from a set of strain images. Several methods for strain tomography have been developed and an overview can be found in \citet{hendriksThesis2020}. Methods for neutron strain tomography take as inputs a large number of strain images requiring many thousands of Bragg-edges to be analysed. Therefore, in order for strain tomography methods to be accurate, it is imperative that the methods used to produce these strain images are accurate and produce reliable estimates of confidence, which the strain tomography methods can take into account \citep{hendriks2019tomographic}.
Additionally, as many of the strain tomography methods use either least-squares \citep{gregg2018tomographic} or an assumption of Gaussian noise \citep{jidling2018probabilistic}, it would be beneficial if the errors given by the methods for Bragg-edge analysis are Gaussian.

The contribution of this paper is twofold. First, it provides a numerical simulation study of the accuracy and reliability of existing methods when applied to data with varying levels of noise. This analysis will show that the more recent approach presented by \citet{ramadhan2018neutron} on average produces smaller errors than the approaches by \citet{santisteban2001time} and \citet{tremsin2016investigation}. However, this is at the cost of very inaccurate quantification of certainty.
Second, a novel non-parametric Bayesian method for estimating strain from measurements of a Bragg-edge is outlined.
This method accurately quantifies the confidence in the produced strain estimates, whilst still giving competitively small errors.
The existing methods and our proposed approach are also demonstrated on two sets of experimental data.

\section{Problem Statement} 
\label{sec:problem_statement}
This paper is concerned with methods for estimating the elastic normal strain $\releps$ from noisy data collected of the transmission $\text{Tr}(\lambda) = \sfrac{I(\lambda)}{I_0(\lambda)}$ over a region containing a single Bragg-edge. 

The data set contains $n$ measurements of the recorded transmission at specified wavelengths $\lambda_i$ which take the form
\begin{equation}\label{eq:noiseModel}
    y_i = \text{Tr}(\lambda_i) + e_i
\end{equation}
where $e$ is additive noise corrupting the measurement. The complete set of measurements will be denoted $y_{1:n} = \{y_1,\dots,y_n\}$.

\begin{figure}[htb]
\centering
\subcaptionbox{\label{fig:ill1}}
{\includegraphics[width=0.45\linewidth]{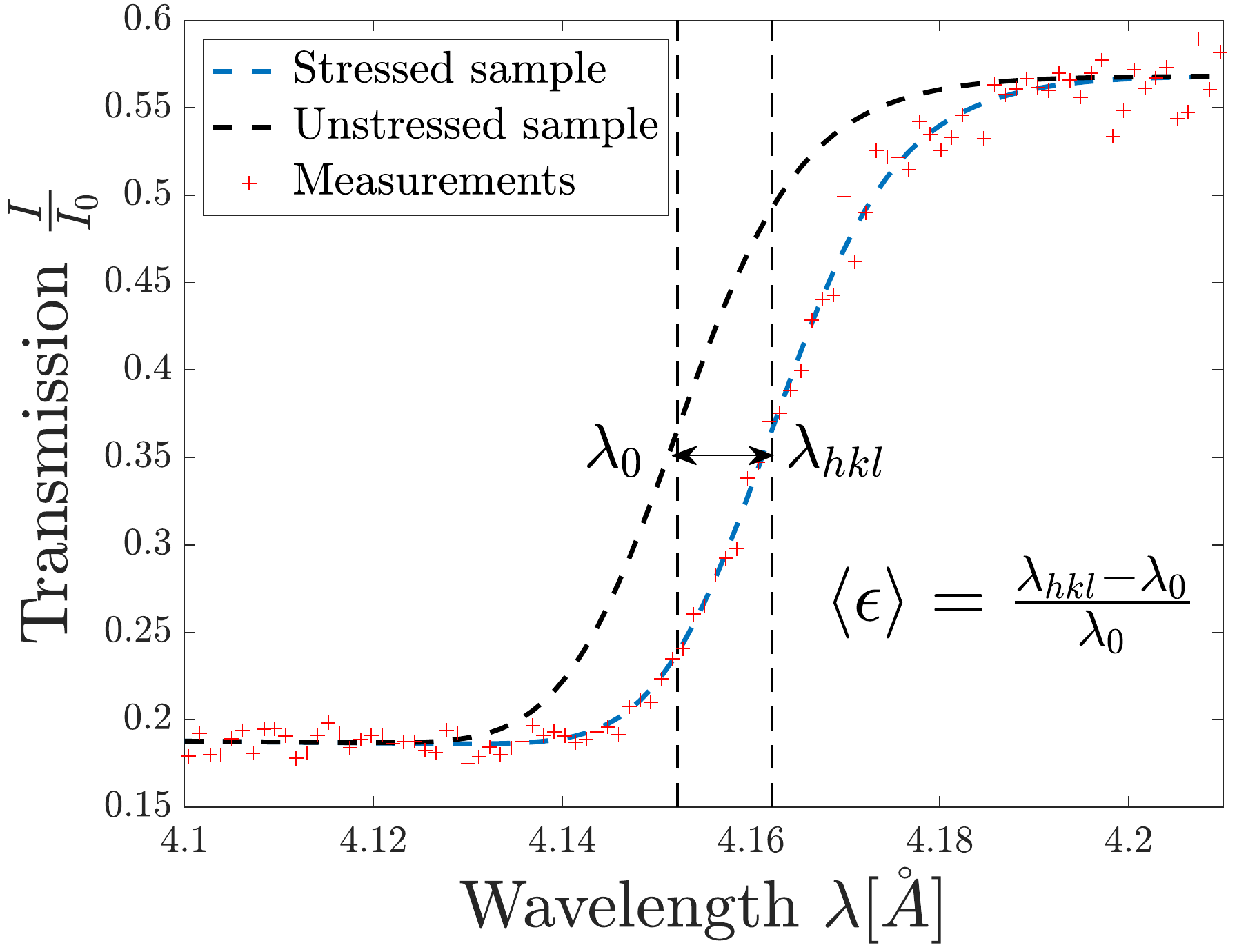}}
\quad
\subcaptionbox{\label{fig:ill2}}
{\includegraphics[width=0.45\linewidth]{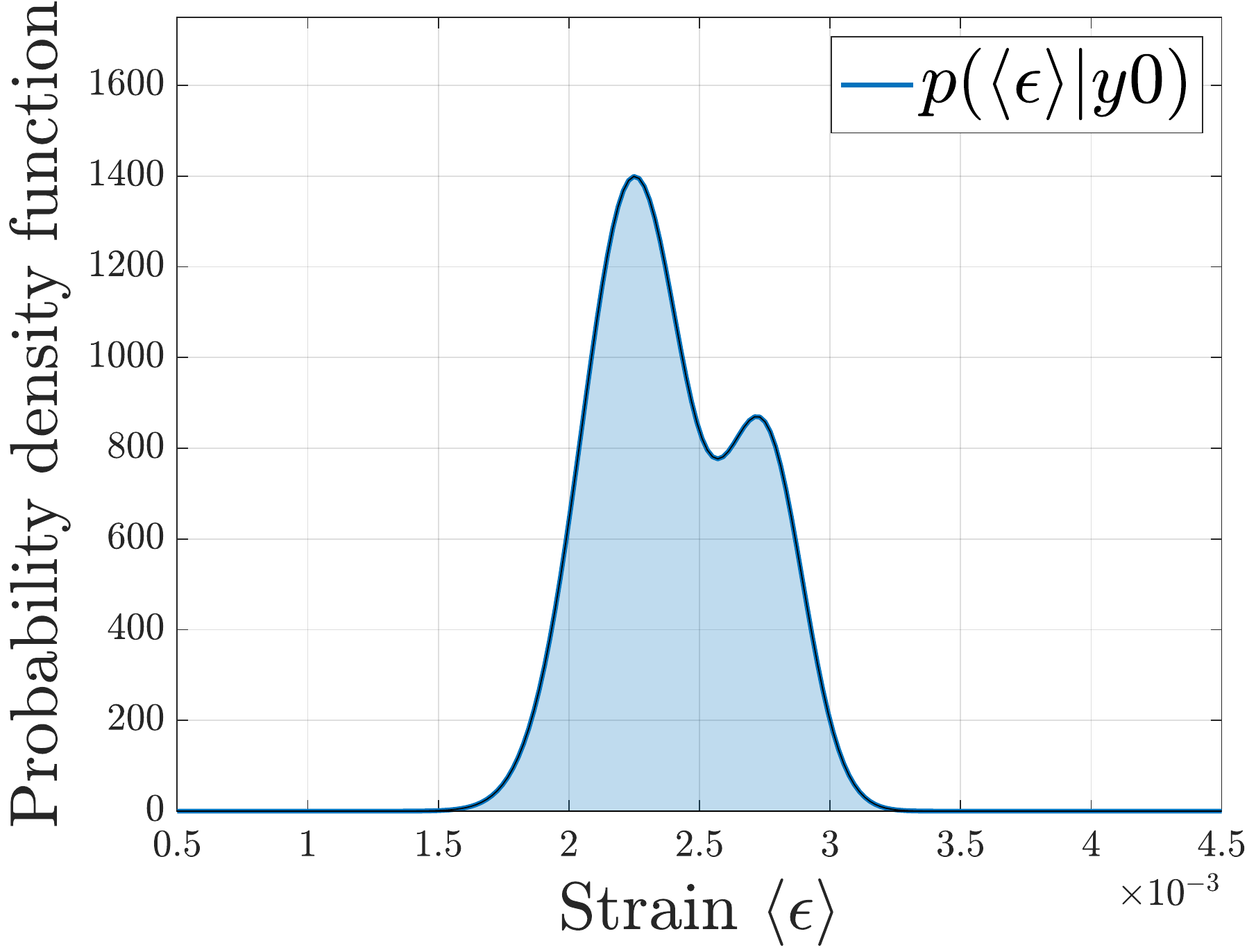}} \\
\caption{An illustrative example of the problem being solved. (\subref{fig:ill1}) Shows a Bragg-edge in a stressed and unstressed sample as well as measurements of the Bragg-edge. Strain can be calculated from the shift in locations of the Bragg-edges $\lambda_{hkl}$ and $\lambda_0$, respectively. (\subref{fig:ill2}) From the measurements of the Bragg-edge we wish to determine a distribution of possible strain values, i.e. the probability density function $p(\langle\epsilon\rangle|y)$.}
\label{fig:illustrative_example}
\end{figure}

Taking a Bayesian viewpoint, we wish to determine the distribution of $\releps$ conditioned on this set of measurements $p(\releps | y_{1:n})$.
An illustration of the problem we are solving is given in Figure~\ref{fig:illustrative_example}.
Since it may not be feasible to determine the full distribution, the methods should provide a reliable estimate of the mean and a suitable measure of certainty, such as confidence interval or standard deviation. 
These estimates of the mean and certainty should be both accurate and robust in the presence of varying levels of noise. That is, given noisy data the method should not produce estimates that are outliers and the provided measure of certainty should reflect the distribution of error between the estimated strain and the true strain.

In the following sections, we will assess the accuracy and robustness of existing methods, described in Section~\ref{sec:existing_approaches}, and compare these to our proposed approach presented in Section~\ref{sec:our_approach}. In order to make this assessment, it is important to determine the distribution of noise corrupting the measurements $p(e|\vartheta)$ and an analysis of this noise is undertaken in Section~\ref{sec:relative_transmission_intensity_noise_analysis}.
The results of this analysis are used in Section~\ref{sec:simulation_demonstration_and_error_analysis}, which undertakes simulated random trials to numerically evaluate the accuracy and robustness of the methods.
In Section~\ref{sec:experimental_data_demonstration} our proposed approach is applied to two sets of experimental data and compared to results from the existing approaches.


\section{Transmission Intensity Noise Analysis} 
\label{sec:relative_transmission_intensity_noise_analysis}
In this section, we undertake an analysis to determine a suitable model for the noise corrupting our measurements of transmission intensity, $p(e \vert \vartheta)$.
For instance, the noise could have a Gaussian distribution, a student's t-distribution, or something more complex and it is useful to know which.
Understanding the way in which noise affects our measurements is necessary for understanding the limiting assumptions of both our method and existing methods and helpful for generating realistic simulation data.

This analysis is performed using two sets of neutron transmission data collected during strain tomography experiments; the first is data from an experiment described in \citet{hendriks2019tomographic} and the second set is from an experiment described Section~\ref{sub:strain_imaging_example}.
These data sets were suitable for this analysis as each contains a large number of measurements --- more than 75 measurements of the transmission spectrum were made of a sample for 2 hours for each experiment --- enabling the distribution of noise to be accurately determined. Due to the short projection times the signal to noise ratio for individual 55 \si{\micro \metre} pixels is generally poor, and, therefore, averaging the data over groups of pixels to create macro pixels was performed to improve this.


In reality, noise enters the measurements through discrete errors in neutron counts at a given pixel on the detector for both the open-beam intensity, $I_0(\lambda)$, and projection intensity, $I(\lambda)$ and manifests itself when the quotient of these two quantities is calculated to produce the \emph{measurement} of transmission. However, we determine that the noise can be modelled as an additive term on the transmission, according to 
\begin{equation}
    y_i = \text{Tr}(\lambda_i) + e_i \tag{\ref{eq:noiseModel} revisited}
\end{equation}


To investigate the properties of $p(e \vert \vartheta)$ we have chosen to obtain samples of $e_i$ by utilising an appropriate model as \emph{ground truth}. Decaying exponential functions, $\exp[ -(a_0 + b_0 \lambda)]$, have been shown to model transmission near a Bragg-edge very well \citep{santisteban2001time}, therefore we have utilised decaying exponential functions fitted to the data using least squares sufficiently far from the edge location where the model fits well and calculating the difference. The exponential model fits well past the last Bragg-edge due to the absence of coherent scattering and fits well before the Bragg-edge provided that the sample does not exhibit significant preferred crystal orientation (texture). These assumptions are valid for our data since we are only considering data immediately left and right of the last Bragg-edge as the data sets used are of samples with minimal grain texture.

\begin{figure}[h]
    \centering
    \begin{minipage}[b]{0.49\linewidth}
            \centering
            \includegraphics[width = \linewidth]{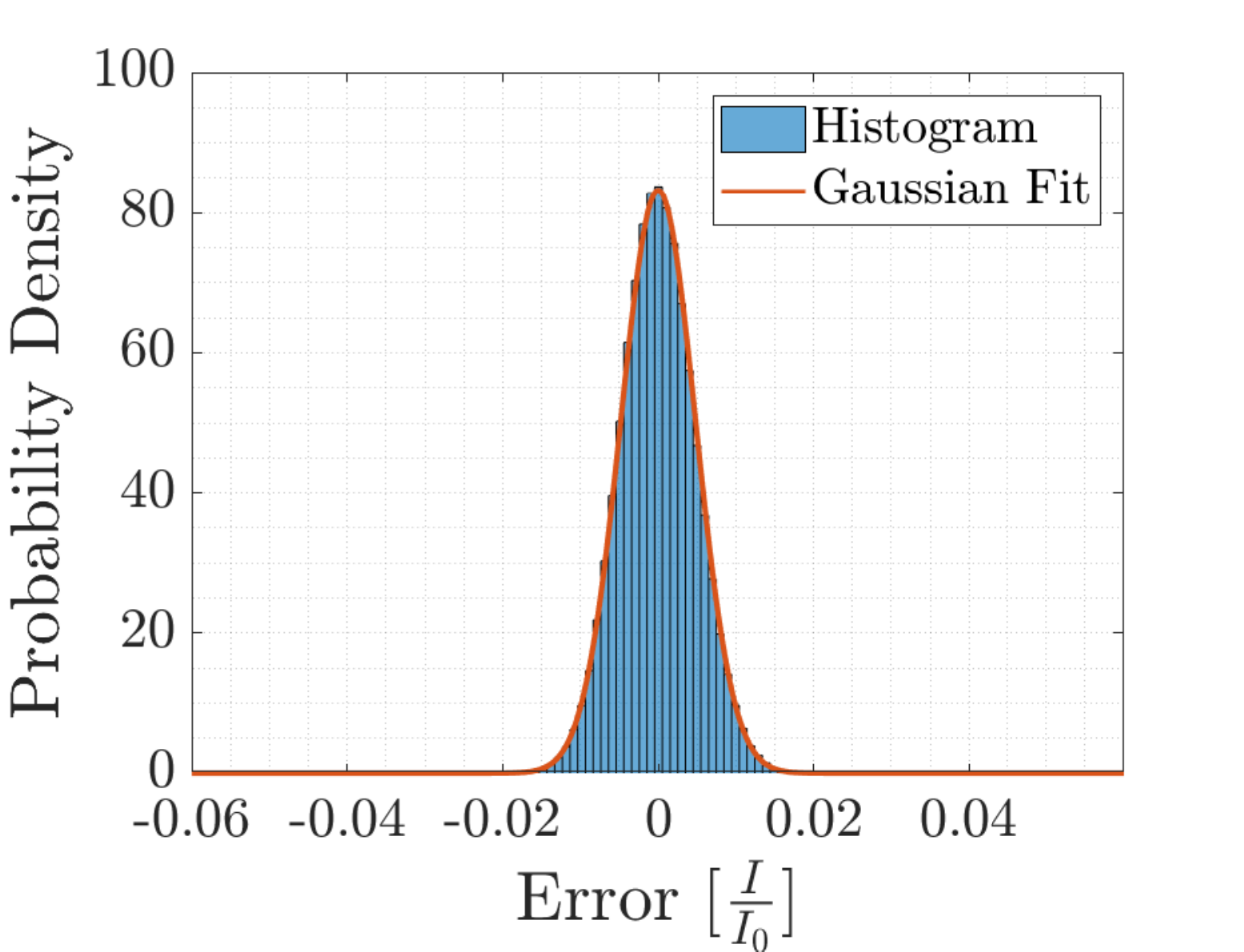}
            \subcaption{\scriptsize$\text{Tr} \in [0.1,0.15], \sigma = 4.79e^{-3}$}
            \label{fig:Hist1}
    \end{minipage}
        \begin{minipage}[b]{0.49\linewidth}
            \centering
            \includegraphics[width = \linewidth]{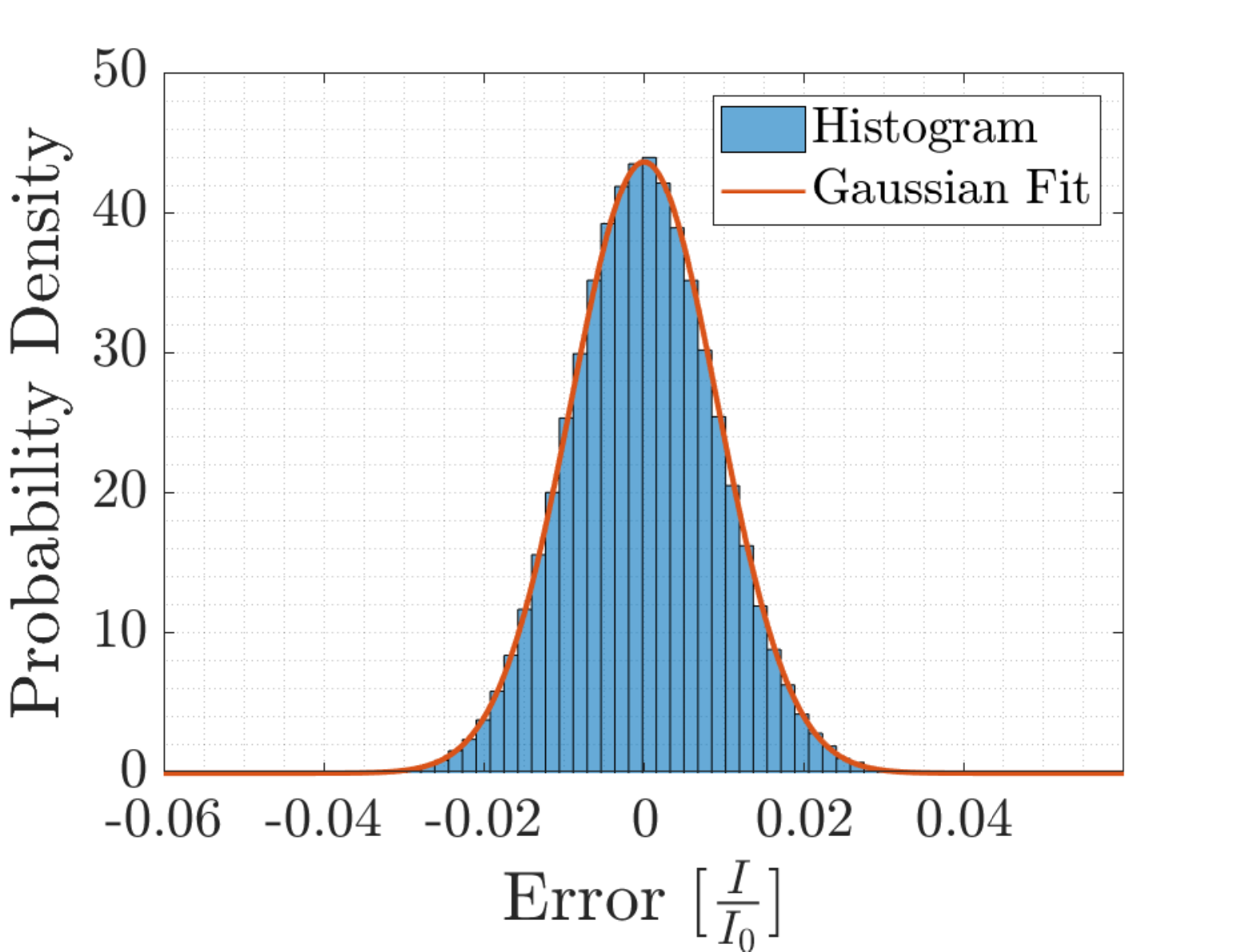}
            \subcaption{\scriptsize$\text{Tr} \in [0.3,0.35], \sigma = 9.14e^{-3}$}
            \label{fig:Hist1}
    \end{minipage}
    \\
        \begin{minipage}[b]{0.49\linewidth}
            \centering
            \includegraphics[width = \linewidth]{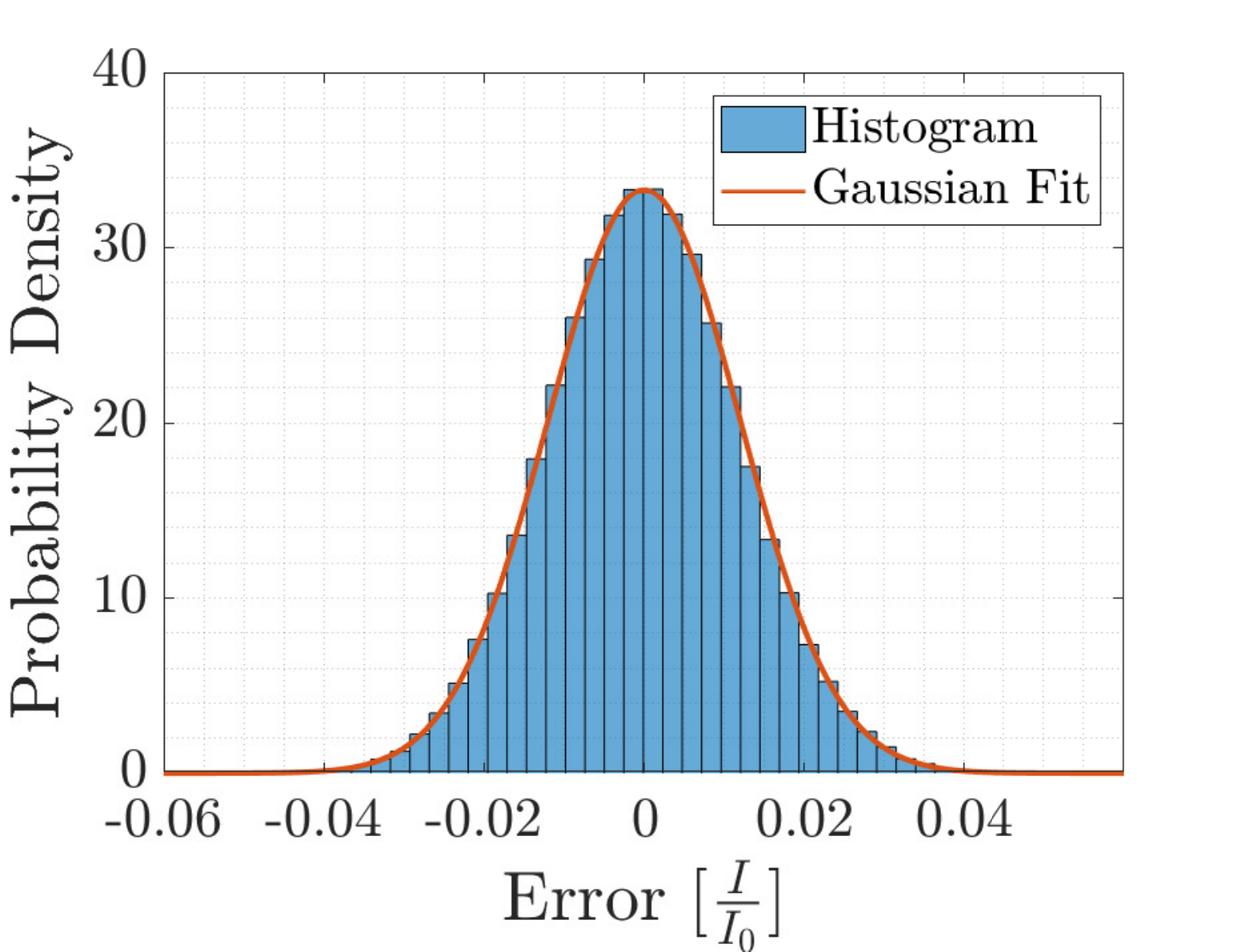}
            \subcaption{\scriptsize$\text{Tr} \in [0.5,0.55], \sigma = 1.20e^{-2}$}
            \label{fig:Hist1}
    \end{minipage}
        \begin{minipage}[b]{0.49\linewidth}
            \centering
            \includegraphics[width = \linewidth]{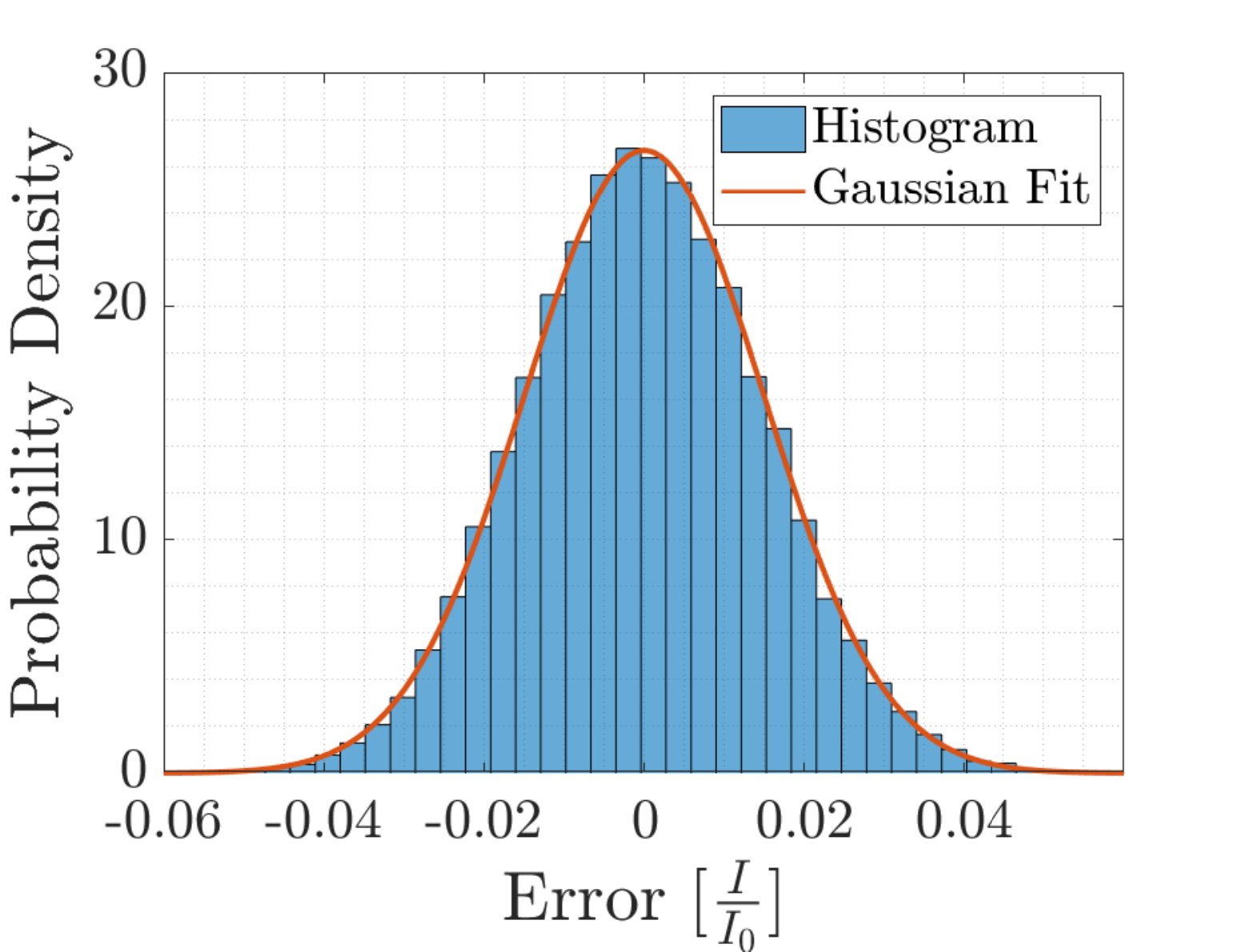}
            \subcaption{\scriptsize$\text{Tr} \in [0.7,0.75], \sigma = 1.50e^{-2}$}
            \label{fig:Hist1}
    \end{minipage}
    \caption{Histograms showing the distribution of the error for different values of Tr from the data set described in Section~\ref{sub:strain_imaging_example}. The data was divided into \emph{bins} according to Tr. A Gaussian probability distribution that has been fit to the data is shown in red.
    }
    \label{fig:histbinTR}
\end{figure}

By inspection of the data, the amount of noise varies with the amount of attenuation. Regions of the data with more attenuation have far less noise than regions of low attenuation, which can be seen in Figure~\ref{fig:GP_Tr_fit}. Therefore by \emph{binning} the data according to ranges of transmission, we can see that the distribution of error conditioned on the true transmission $p(e_i \vert \operatorname{Tr}(\lambda_i))$, is very close to a zero-mean Gaussian, shown in Figure~\ref{fig:histbinTR}.

\begin{figure}[htb]
    \centering
    \begin{minipage}[b]{0.49\linewidth}
        \centering
        \includegraphics[width = 1\linewidth]{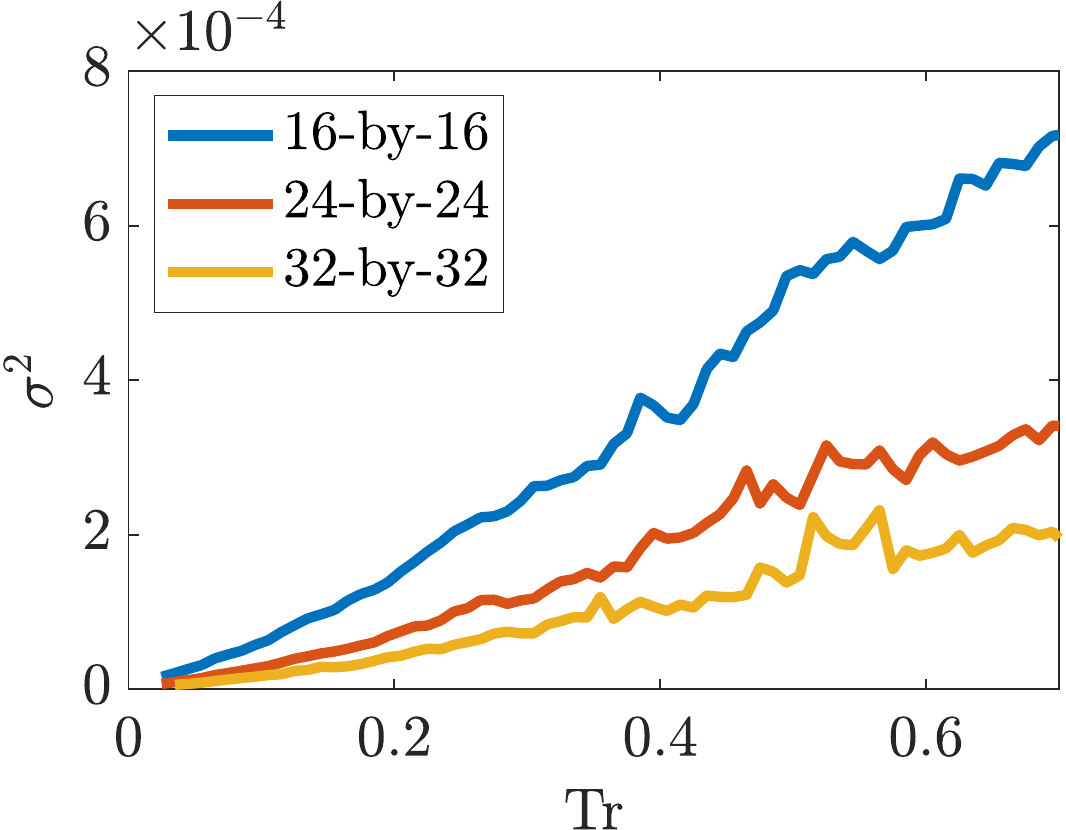}
        \subcaption{~}
        \label{fig:JPARC2019trend}
    \end{minipage}
        \begin{minipage}[b]{0.49\linewidth}
        \centering
        \includegraphics[width = 1\linewidth]{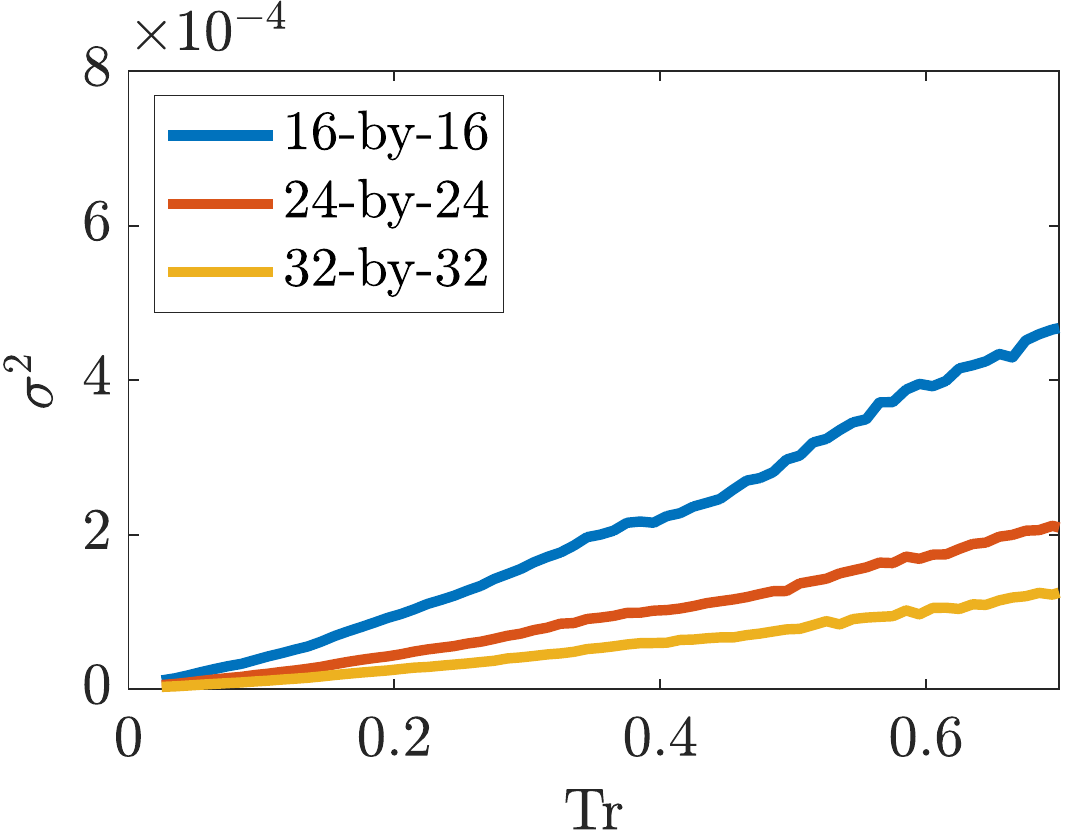}
        \subcaption{~}
        \label{fig:JPARC2020trend}
    \end{minipage}
    \caption{Relationship between variance and attenuation for two experimental data sets. (a) presents the trend for the data used by \citet{hendriks2019tomographic}. (b) presents the same trend for a data set obtained by the authors, presented in Section~\ref{sub:strain_imaging_example}. The relationship is shown for 3 macro pixels sizes.
    }
    \label{fig:sigVsTr}
\end{figure}

Following on from the result in Figure~\ref{fig:histbinTR}, we can use a Gaussian distribution to model the noise and establish a relationship between transmission ratio and the variance of our Gaussian model, this relationship is shown in Figure~\ref{fig:sigVsTr} for three macro-pixel sizes. From this we can establish that variance appears to be a linear function of the transmission ratio, we can therefore write
\begin{equation}
p\left(y_i \vert \text{Tr}(\lambda_i)\right) \sim \mathcal{N}\left( \text{Tr}(\lambda_i),\sigma_i^2\right)
\end{equation}
where,
\begin{equation}
\sigma_i^2 = a + b\left[\text{Tr}(\lambda_i)\right].
\end{equation}
Importantly, the form of the distribution is not affected by how many pixels are combined to create a measurement, only that the variance of the noise decreases as more pixels are included in the measurement.


Several pertinent conclusions can be drawn from these results which have implications for both our proposed method and existing ones. As the distributions are incredibly close to Gaussian, we can conclude that that the use of a Gaussian likelihood model is appropriate. This also validates the choice of a least-squares cost function proposed by some existing approaches described in Section~\ref{sec:existing_approaches}. Additionally, we can conclude that due to the absence of a bias, the use of the exponential functions to locally model attenuation far from the edge is appropriate.
\section{Existing Approaches} 
\label{sec:existing_approaches}
This section provides a brief overview of existing approaches.

The predominant approaches used for determining strain images from neutron Bragg-edge data are presented by \citet{santisteban2001time,santisteban02b}, and \citet{tremsin2016investigation,tremsin12}. Both of these approaches fit a parametric model to the recorded transmission to determine $\lambda_{hkl}$ from which strain is determined using Equation~\ref{eq:relative_strain}. 
Recently, an alternative approach that uses cross-correlation to estimate $\Delta_{hkl} = \lambda_{hkl}-\lambda_0$ has been presented by \citet{ramadhan2019characterization,ramadhan2018neutron}. For readability, we will refer to these methods as the Santisteban method, the Tremsin method, and the cross-correlation method, respectively. 

There are also methods available that analyse the full spectrum (containing multiple Bragg-edges) \citep{vogel2000tof,sato2011rietveld}. 
Ideally, such a full spectrum analysis is used for analysis of micro-structure in the sample, including strain.
Such analysis has only been demonstrated for a small number of measured spectra as it requires substantial computing power to fit a large number of parameters and some \emph{a priori} information about the micro-structure. 
Hence, the analysis of the entire spectrum for strain imaging and strain tomography seems to be impractical at the present time where hundreds or thousands of transmission spectrum may need to be analysed. 
Additionally, single edge fitting is implemented not only because of acceptable computing time, but because different Bragg edges can have a different response to a specific stress induced on the sample. 

The Santisteban method models the recorded transmission by
\small
\begin{equation}\label{eq:santistban_model}
\begin{split}
    \text{Tr}(\lambda) =& B(\lambda,\lambda_{hkl},\sigma_B,\tau)\exp(-a_0-b_0\lambda)\\ &+ (1-B(\lambda,\lambda_{hkl},\sigma_B,\tau))\exp(-a_0-b_0\lambda)\\&\hspace{20mm}\exp( - a_{hkl} - b_{hkl}\lambda), 
\end{split}
\end{equation}
\normalsize
where the exponential terms provide good models for the data either side of the edge and the edge shape $B$ is chosen as the integral of the Kropff model \citep{kropff1982bragg} given by
\small
\begin{equation}\label{eq:kropff_model}
\begin{split}
    B(\lambda,\lambda_{hkl},\sigma_B,\tau) =& \frac{1}{2}\Bigg[\text{erfc}\left(-\frac{\lambda-\lambda_{hkl}}{\sqrt{2}\sigma_B}\right)\\ &- \exp\left(-\frac{\lambda-\lambda_{hkl}}{\tau}+\frac{\sigma_B^2}{2\tau^2}\right)\\&\hspace{9mm}\text{erfc}\left(-\frac{\lambda-\lambda_{hkl}}{\sqrt{2}\sigma_B}+\frac{\sigma_B}{\tau}\right)\Bigg].
\end{split}
\end{equation}
\normalsize
The model is fit to the data using a least-squares method, such as the Matlab function \texttt{lsqcurvefit}. To help avoid local minima that exist due to the non-linear nature of the model, \citet{santisteban2001time} suggest fitting the model in three stages. First, $a_0$ and $b_0$ are estimated by fitting $\exp(-a_0-b_0\lambda)$ to the far side of the edge where $B(\lambda,\lambda_{hkl},\sigma_B,\tau)=1$. Then, $a_{hkl}$ and $b_{hkl}$ are estimated by fitting $\exp(-a_0-b_0\lambda-a_{hkl}-b_{hkl}\lambda)$ to the far left of the edge where $B(\lambda,\lambda_{hkl},\sigma_B,\tau)=0$. Finally, the values of $\lambda_{hkl}$, $\sigma$, and $\tau$ are estimated by fitting the full model to data spanning the Bragg-edge. In each stage, the previously estimated parameters are held constant.

Additionally, \citet{santisteban2001time} mentions that an alternative, more complex, model for the Bragg-edge presented by \citet{vogel2000tof} can be used. This models the Bragg-edge by the integral of a Gaussian convoluted with two back-to-back exponentials with different rise and decay rates;
\small
\begin{equation}\label{eq:vogel_model}
    B(\lambda,p)\hspace{-1mm} =\hspace{-1mm}\frac{1}{2}\text{erfc}(w)  - \frac{\beta\exp(u)\text{erfc}(y)\hspace{-1mm} -\hspace{-1mm} \alpha\exp(v)\text{erfc}(z)}{2(\alpha+\beta)} ,
\end{equation}
\normalsize
where $p = \{\lambda_{hkl},\sigma_B,\alpha,\beta\}$ and
\begin{equation}
\begin{split}
    \delta =& \lambda_{hkl}-\lambda, \quad 
    w = \frac{\delta}{\sqrt{2}\sigma_B}, \\
    u =& \frac{\alpha}{2}(\alpha\sigma_B^2 + 2\delta), \quad
    v = \frac{\beta}{2}(\beta\sigma_B^2 - 2\delta), \\
    y =& \frac{\alpha\sigma_B^2+\delta}{\sqrt{2}\sigma_B}, \quad
    z = \frac{\beta\sigma_B^2-\delta}{\sqrt{2}\sigma_B}.
\end{split}
\end{equation}
In the authors' experience, this more complex model provides a better fit for some edge shapes. However, it becomes more difficult to avoid local minima while fitting. 

The Tremsin method fits a five parameter model to the region around the Bragg-edge. This method does not use exponential models for the attenuation either side of the edge. Instead, it adds additional parameters to the Kropff Bragg-edge model given in Equation~\ref{eq:kropff_model}: edge height, $h$, and base height $b$. This gives the model as
\small
\begin{equation}
    B(\lambda,\lambda_{hkl},\sigma_B,\tau,b,h) = h*B(\lambda,\lambda_{hkl},\sigma_B,\tau,b,h) + b
\end{equation}
\normalsize
This simple model is easy to implement and fit using non-linear least squares. However, this model has been found to less adequately describe the slope trends on the left-hand side of the edge for certain data sets \citep{ramadhan2018neutron}. Despite this, in the authors experience it can give good results provided that it is fit over a region of the transmission tightly cropped around the Bragg-edge.

Since both the Santisteban and Tremsin method fit a parametric model to the edge shape, the performance of both of these methods may suffer if the Bragg-edge to be fit is of a shape that cannot be adequately represented by these models. This could occur, for instance, in samples with significant preferred crystal orientation (texture), which distorts the edge shape.
This was a significant motivating factor behind the development of the cross-correlation method by \citet{ramadhan2018neutron} and also motivates our use of a non-parametric edge shape in Section~\ref{sec:our_approach}.

The cross-correlation method proceeds as follows. First, smoothed numerical derivatives of the recorded transmission for both the stressed and stress-free sample are computed. Second, cross-correlation is performed on the two derivatives which determines the correlation between the edge shapes as a function of displacement $\Delta=\lambda-\lambda_0$. third, a pseudo Voigt function is fit to the peak-shaped correlation curve;
\small
\begin{equation*}
\begin{split}
    V(\Delta,\Delta_{hkl},A,\mu,w_l,w_g) =& y_0 + A\Bigg[ \frac{\mu  2\pi w_l}{4(x-\Delta)^2+w_l^2}  \\
    &+(1-\mu)\frac{\sqrt{4\log(2)}}{\sqrt{\pi}w_g} \\
    &\exp\Bigg(-\frac{4\log(2)(\Delta-\Delta_{hkl})^2}{w_g^2}\Bigg)\Bigg].
\end{split}
\end{equation*}
\normalsize
Once $\Delta_{hkl}$ is estimated, an estimate of strain can be determined given knowledge of $\lambda_{0}$ according to $\releps=\sfrac{\Delta_{hkl}}{\lambda_0}$.

The use of cross-correlation between the $\lambda_0$ and $\lambda_{hkl}$ data allows this method to be applied to a wide variety of Bragg-edge shapes. However, it is unclear if the $\lambda_0$ Bragg-edge and the $\lambda_{hkl}$ Bragg-edge having significantly different shapes would bias the strain estimates.
Additionally, since differentiation magnifies noise present in data, a method for producing smoothed derivatives, such as the Savitzky-Golay method \citep{gorry1990general} is required. To achieve good results for different noise levels this requires manual tuning of the fit window and polynomial order parameters.

Each of these methods requires a parametric model to be fit. In \citet{santisteban2001time} and \citet{tremsin2016investigation} it is specified that non-linear least squares is used. 
Given that the analysis in Section~\ref{sec:relative_transmission_intensity_noise_analysis} determined the noise to have a Gaussian distribution we can conclude that the choice of a least-squares fitting method is appropriate. This is because least-squares is equivalent to finding a maximum likelihood estimate with a Gaussian noise model \citep{hendriksThesis2020}.
As such, these methods can be interpreted as providing a maximum likelihood estimate of strain;
\begin{equation}
    \releps_{ML} = \argmax_{\releps} p(y_{1:n},\releps),
\end{equation}
which is equivalent to the maximum \textit{a posteriori} estimate given a uniform prior for $p(\releps)$.

Having obtained an estimate using least squares, it is also possible quantify the certainty of this estimate. 
A typical approach is to determine the Fisher information matrix, which can be estimated around the solution using Taylor's theorem, from which confidence intervals or covariance can be calculated \citep{geyer2007fisher}.
However, due to the non-linear nature of the models this estimate of the confidence interval may not always be accurate, particularly if the true conditional distribution $p(\releps | y_{1:n})$ is non-Gaussian or in extreme cases multi-modal.
Additionally, as with most non-linear optimisation problems, there is no guarantee that a global minima is found. If instead a local minima is found, then both the estimate of strain and its confidence interval could be wrong.


\section{Proposed Bayesian Approach} 
\label{sec:our_approach}
In this section, we propose a non-parametric Bayesian approach for estimating strain from time-of-flight neutron transmission Bragg-edge measurements.
The approach models the Bragg-edge shape using a Gaussian process. This non-parametric model is beneficial as it is capable of fitting a wide range of edge shapes. However, it becomes more challenging to determine strain from these fits as there is no longer a model parameter that represents $\lambda_{hkl}$. This is overcome, by using the maximum of the derivative which can be exactly computed for a Gaussian process.

Our approach is presented as follows.
A brief introduction to Gaussian process regression is presented in Section~\ref{sub:an_introduction_to_gaussian_processes}.
Section~\ref{sub:gaussian_process_bragg_edge_fitting} describes a procedure to fit the transmission data using a Gaussian process model for the edge shape.
A method is outlined in Section~\ref{sub:strain_measurements} for determining the strain from this non-parametric estimate of the edge shape.
Lastly, Gaussian processes have a number of hyperparameters; for example, the length-scale which controls the characteristic smoothness of the function, and Section~\ref{sub:hyperparameter_optimisation} outlines how to optimise these parameters.

\subsection{An Introduction to Gaussian Processes Regression} 
\label{sub:an_introduction_to_gaussian_processes}
This section provides a brief introduction to Gaussian process regression, a more thorough discourse is given by \citet{rasmussen2006gaussian}.

Gaussian process regression is a non-parametric Bayesian method for fitting spatially correlated functions.
A Gaussian process (GP) is a non-parametric Gaussian distribution of functions that is fully defined by a mean function $m:\mathbb{R}\to\mathbb{R}$ and covariance function $k:\mathbb{R}\times\mathbb{R}\to\mathbb{R}$.
For clarity, we restrict the input space to be scalar.

The covariance function $k$ determines the correlation between any two function values $f(x)$ and $f(x')$ given by $k(x,x')$. 
As such, the choice of covariance function determines the characteristics of functions belonging to a Gaussian process, such as the degree of smoothness and differentiability of these functions.
Two common covariance functions are the squared-exponential, $k_{SE}$, and Mat\'{e}rn, $k_M$, given by
\begin{equation}
\begin{split}
    k_{SE}(x,x') &= \sigma_f\exp\left(-\frac{1}{2l^2}(x-x')^2\right), \\
    k_{M}(x,x') &= \sigma_f\frac{2^{1-\nu}}{\Gamma(\nu)}\left(\frac{\sqrt{2\nu}r}{l}\right)^\nu I_\alpha\left(\frac{\sqrt{2\nu}r}{l}\right),
\end{split}
\end{equation}
where $r=|x-x'|$ is the distance between input locations, $l$ is the length-scale, $\sigma_f$ is the prior variance, $\nu$ is the degrees of freedom of the Mat\'{e}rn covariance function, and $I_\alpha$ is the modified Bessel function \citep{abramowitz1964handbook}.
With both these covariance functions, the correlation between two function values is proportional to the distance between their input locations scaled by the length-scale. Hence, a larger length-scale will impose a greater degree of smoothness on the functions. 
The length-scale and prior variance are commonly referred to as hyperparameters and their selection and the selection of covariance function is discussed in Section~\ref{sub:hyperparameter_optimisation}.

For a function belonging to a GP distribution, any finite set of function values at locations have a multivariate Gaussian distribution. 
Hence, an unknown function value $f_*$ at $x_*$ and a measurement of the function 
\begin{equation}\label{eq:GP_measurement}
    y_i=A_if(x_i)+e,
\end{equation}
where $A$ is a linear mapping and e is Gaussian noise with standard deviation $\sigma$, are \textit{a priori} multivariate Gaussian\footnote{The notation $\mathcal{N}(\mu,\Sigma)$ is used to denote a multivariate Gaussian with mean $\mu$ and covariance $\Sigma$.};
\small
\begin{equation*}\setlength\arraycolsep{1pt} 
\pmat{y_i \\ f_*} \sim \mathcal{N}\left(\pmat{A_i m(x_i)  \\ m(x_*)
},\pmat{A_ik(x_i,x_i)A_i^{\Transp}\hspace{-1.1mm} +\hspace{-0.6mm} \sigma^2 &  A_ik(x_i,x_*) \\
k(x_*,x_i)A_i^{\Transp} & k(x_*,x_*) 
}\right)
\end{equation*}
\normalsize
Without loss of generality, we can assume the \textit{a priori} mean function to be zero \citep{rasmussen2006gaussian}.
Given that we know the measurement value, we can update our knowledge of $f_*$ using standard Gaussian conditioning
\begin{equation}\label{eq:GP_posterior}
\begin{split}
    f_*| y_i \sim \mathcal{N}\left(\mu_{f_*|y_i},\Sigma_{f_*|y_i}\right)
\end{split}
\end{equation}
where
\begin{equation*}
\begin{split}
    \mu_{f_*|y_i} &= k(x_*,x_i)A_i^{\Transp}(A_ik(x_i,x_i)A_i^{\Transp}+\sigma^2)^{-1}y_i \\
    \Sigma_{f_*|y_i} &= k(x_*,x_*) - k(x*,x_i)A_i^{\Transp}\\&\hspace{10mm}(A_ik(x_i,x_i)A_i^{\Transp}+\sigma^2)^{-1}A_ik(x_i,x_*)
\end{split}
\end{equation*}

Note that this is trivially extended to conditioning a set of function values on a set of measurements.
This means that Gaussian process regression can easily be used to condition an estimate of the unknown functions over a range of input locations based on a set of measurements.

\subsection{Gaussian process Bragg-edge fitting} 
\label{sub:gaussian_process_bragg_edge_fitting}
Here, a method for fitting Bragg-edges using Gaussian process regression is presented. 
To do this, the transmission intensity $\text{Tr}(\lambda) = \sfrac{I(\lambda)}{I_0(\lambda)}$ is modelled similarly to the approach in \citet{santisteban2001time}, using two decaying exponentials and an edge shape function, see Equation~\eqref{eq:santistban_model}.

Our approach differs in that we use a Gaussian process to model the edge shape $B(\lambda)$ rather than a parametric model.
Rearranging Equation~\eqref{eq:santistban_model} we can write
\begin{equation}
\begin{split}
    \text{Tr}(\lambda) - \gamma_1(\lambda) = (\gamma_2(\lambda) - \gamma_1(\lambda))B(\lambda),
\end{split}
\end{equation}
where $\gamma_1(\lambda) = \exp(-(a_0+b_0\lambda+a_{hkl}+b_{hkl}\lambda))$ and $\gamma_2(\lambda) = \exp(-(a_0+b_0\lambda))$.
If we now let $\bar{y}_i = y_i - \gamma_1(\lambda_i)$ and $A(\lambda) = (\gamma_2(\lambda_i)-\gamma_1(\lambda_i))$ then our measurements are of the same form as Equation~\ref{eq:GP_measurement}, and Equation~\ref{eq:GP_posterior} can be applied to estimate the unknown function $B$.

The Bragg-edge can be fit using the following procedure
\begin{enumerate}
    \item The values of $a_0$ and $b_0$ are determined by fitting $\exp(-(a_0 + b_0\lambda))$ to the recorded transmission to the far right of the edge.
    \item Then, $a_{hkl}$ and $b_{hkl}$ are determined by fitting $\exp(-(a_0 + b_0\lambda - a_{hkl}-b_{hkl}\lambda))$ to the recorded transmission to the far left of the edge.
    \item Lastly, Gaussian process regression is used to estimate $B(\lambda)$.
\end{enumerate}

\begin{figure}[htb]
    \centering
    \includegraphics[trim={5 0 35 20},clip,width=1.0\linewidth]{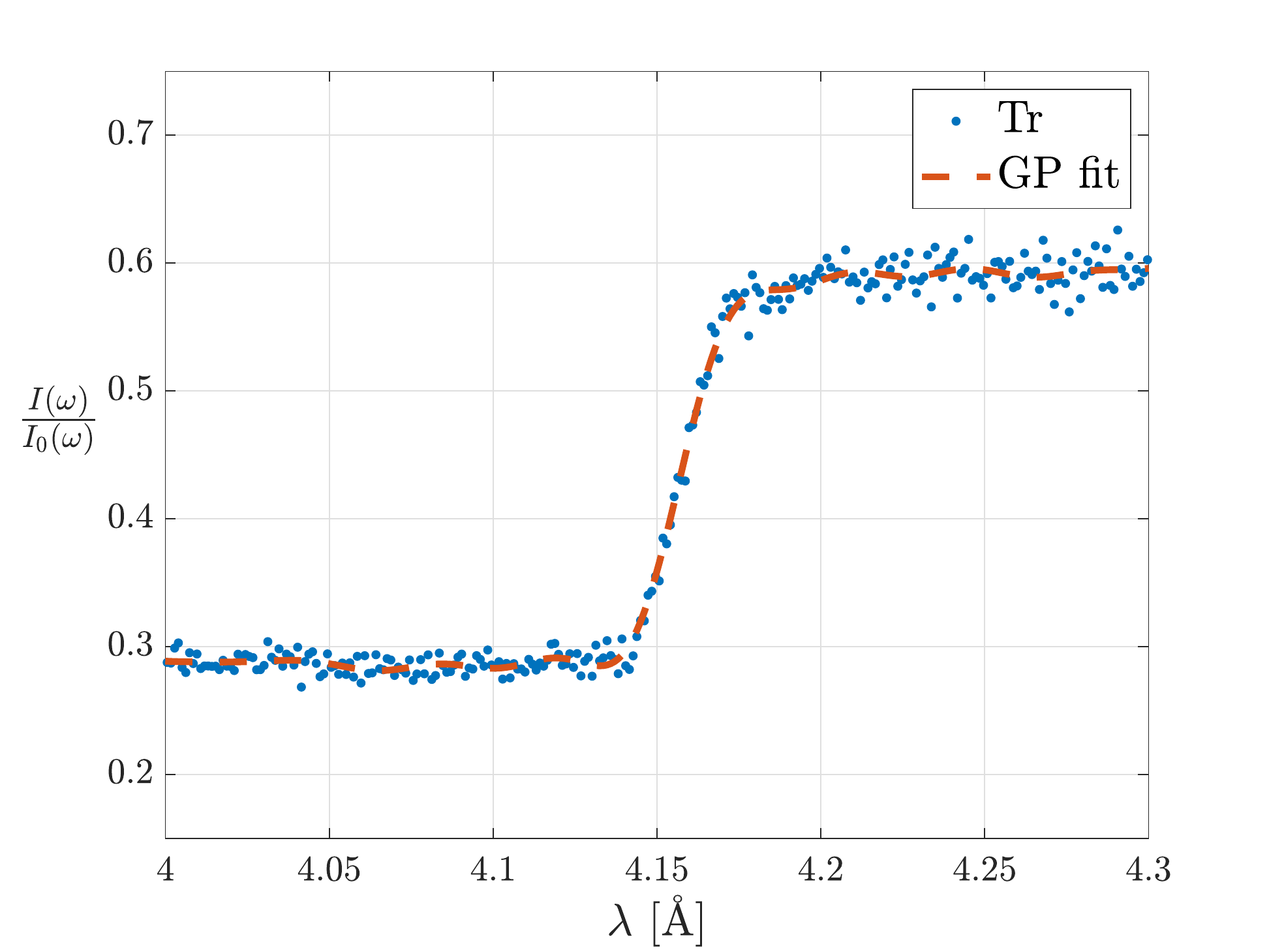}
    \caption{An example of fitting the transmission according to Equation~\eqref{eq:santistban_model} with the edge shape modelled by a Gaussian process using a Mat\'{e}n covariance function with $\nu=\sfrac{3}{2}$.}
    \label{fig:GP_Tr_fit}
\end{figure}

An example of using this approach to fit a recorded transmission belonging to the data set described in Section~\ref{sub:strain_imaging_example} is shown in Figure~\ref{fig:GP_Tr_fit}.

\subsection{Strain measurements} 
\label{sub:strain_measurements}
Having fit the transmission data using a Gaussian process model for the edge shape we now wish to determine the distribution of strain values, $p(\langle\epsilon\rangle | Y)$.
This distribution will provide us with both the expected value and the confidence in this value.

Using a non-parametric model for the edge shape has the advantage that it is not constrained to a particular shape. 
However, it presents the challenge that this edge shape model does not have a $d_{hkl}$ parameter and so Equation~\ref{eq:relative_strain} cannot be directly applied to determine the strain.
Therefore, our approach determines the strain from the relative shift in the maximum gradient of the edge shape;
\begin{equation}
    \langle \epsilon \rangle = \frac{\zeta - \zeta_0}{\zeta_0}
\end{equation}
where
\begin{equation}\label{eq:lambda_equation}
    \zeta = \argmax_{\bar{\lambda}} \frac{\partial}{\partial\lambda}B(\lambda)\Big|_{\lambda=\bar{\lambda}}.
\end{equation}
That is, $\zeta$ is the wavelength for which the gradient of the edge shape is maximised. Similarly, $\zeta_0$ is computed in the same manner from the transmission recorded of a stress-free sample.

\begin{figure}[htb]
    \centering
    \includegraphics[trim={0 0 0 0},clip,width=1.0\linewidth]{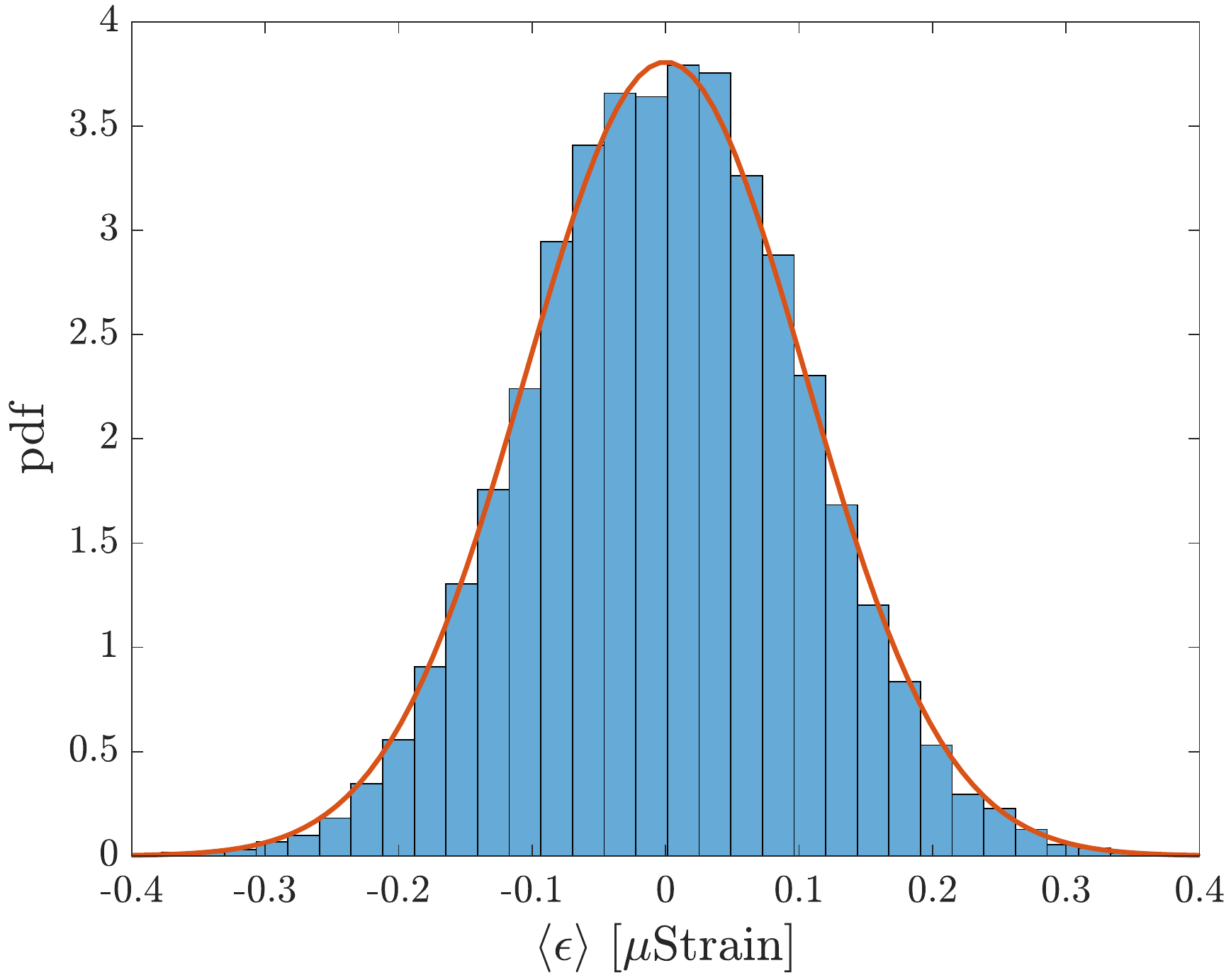}
    \caption{A histogram showing the Monte Carlo approximation of $p(\langle\epsilon\rangle|y_{1:n})$. The samples are computed according to Equation~\eqref{eq:eps_samples} and correspond to the Bragg-edge fit shown in Figure~\ref{fig:GP_Tr_fit}. The Gaussian probability density function with mean and variance given by Equation~\ref{eq:sample_mean_var} is also shown in red.}
    \label{fig:eps_hist}
\end{figure}

An advantage of using Gaussian process regression to model the edge shape is that the required derivative can be computed directly, avoiding the use of numerical differentiation.
This is possible as differentiation is a linear operator and Gaussian processes are closed under linear operators \citep{rasmussen2006gaussian,wahlstrom2015modeling}. 
As a consequence, the gradient of the edge shape, $g=\frac{\partial B(\lambda)}{\partial\lambda}$, at $\lambda_*$ and a measurement, $\bar{y}_i=Tr(\lambda_i)-\gamma_1(\lambda_i)$, at $\lambda_i$ are \textit{a priori} multivariate Gaussian;
\small
\begin{equation*}\setlength\arraycolsep{1pt} 
\begin{split}
\pmat{\bar{y}_i \\ g_*} &\sim  \mathcal{N} \left(\pmat{0 \\ 0
},\pmat{Ak(x_i,x_i)A^{\Transp} + \sigma^2 & k_*^{\Transp} \\ k_* & k_{**}}
\right), \\
k_* &=  \frac{\partial}{\partial \lambda}k(\lambda,\lambda_i)A(\lambda_i)^{\Transp}\Big|_{\lambda=\lambda_*}, \\
k_{**} &= \frac{\partial^2}{\partial \lambda\partial\lambda'}k(\lambda,\lambda')\Big|_{\substack{\lambda=\lambda_*\\\lambda'=\lambda_*}}, \\
\end{split}
\end{equation*}
\normalsize
where the prior mean was assumed zero.
Therefore the gradient of the edge shape conditioned on a measurement, $p(g|\bar{y}_i)$, will be Gaussian with mean $\mu_{g|\bar{y}_i}$ and covariance $\Sigma_{g|\bar{y}_i}$ according to
\begin{equation}
\begin{split}
    \mu_{g|\bar{y}_i} &= k_*(Ak(x_i,x_i)A^{\Transp} + \sigma^2 )^{-1}\bar{y}_i, \\
    \Sigma_{g|\bar{y}_i} &= k_{**} - k_*(Ak(x_i,x_i)A^{\Transp} + \sigma^2 )^{-1}k_*^\Transp.
\end{split}
\end{equation}
This is easily extended to compute $p(g|\bar{y}_{1:n})$ from which $p(\zeta|\bar{y}_{1:n})$ can be determined by solving
\begin{equation}
    p(\zeta | \bar{y}_{1:n}) = \argmax_{\bar{\lambda}}g(\lambda)p(g(\lambda);g|\bar{y}_{1:n}).
\end{equation}
Since this is analytically intractable Monte Carlo sampling can be used to draw samples from $p(\zeta| \bar{y}_{1:n})$, $p(\zeta_0 | \bar{y}_{1:n})$ and $p(\langle\epsilon\rangle|y_{1:n})$ according to;
\begin{equation}\label{eq:eps_samples}
\begin{split}
    \zeta^{(i)} &= \argmax_{\bar\lambda} \mathcal{G}^{(i)}(\lambda), \quad \mathcal{G}^{(i)} \sim \mathcal{N}(\mu_{g|\bar{y}_{1:n}} | \Sigma_{g|\bar{y}_{1:n}}),
     \\
     \zeta_0^{(i)} &= \argmax_{\bar\lambda} \mathcal{G}_0^{(i)}(\lambda), \quad \mathcal{G}_0^{(i)} \sim \mathcal{N}(\mu_{g_0|\bar{y}_{1:n}} | \Sigma_{g_0|\bar{y}_{1:n}}),
     \\
    \langle\epsilon\rangle^{(i)} &= \frac{\zeta^(i)-\zeta_0^{(i)}}{\zeta_0^{(i)}}
\end{split}
\end{equation}
The samples of the edge shape gradient $\mathcal{G}^{(i)}$ and stress-free edge shape gradient $\mathcal{G}_0^{(i)}$ are easy to compute given that they have Gaussian distributions. Given $N$ samples $\langle\epsilon\rangle$ the sample mean and variance of $p(\langle\epsilon\rangle|\bar{y}_{1:n})$ is given by
\begin{equation}\label{eq:sample_mean_var}
\begin{split}
    \mu_{\langle\epsilon\rangle|y_{1:n}} &= \frac{1}{N}\sum_i^N \langle\epsilon\rangle^{(i)}, \\
    \Sigma_{\langle\epsilon\rangle|y_{1:n}} &= \frac{1}{N}\sum_i^N(\langle\epsilon\rangle^{(i)} - \mu_{\langle\epsilon\rangle|\bar{y}_{1:n}})^2.
\end{split}
\end{equation}
Figure~\ref{fig:eps_hist} shows a histogram of $\langle\epsilon\rangle^{(i)}$ samples corresponding to the Bragg-edge data shown in Figure~\ref{fig:GP_Tr_fit}. The Gaussian probability density function (pdf) with mean and variance given by Equation~\eqref{eq:sample_mean_var} is also plotted.
The Gaussian pdf very closely matches the histogram indicating that, for this data, the mean and variance are sufficient to describe the distribution $p(\releps|y_{1:n})$.

Before proceeding, a few pertinent remarks can be made. First, $\lambda$ should not be used as an estimate of the lattice spacing $\lambda_{hkl}$. This is apparent by considering the Kropff model \citep{kropff1982bragg} for the instrument resolution function with non-zero asymmetry. In this case, the value for $\lambda_{hkl}$ will be offset from the peak value of the derivative. This offset is not a problem when calculating strain provided that the asymmetry is relatively consistent between the stress-free Bragg-edge and the Bragg-edge used to calculate strain since the offset will cancel.

Second, although the Monte Carlo approximation of $p(\releps|\bar{y}_{1:n})$ shown in Figure~\ref{fig:eps_hist} is very close to Gaussian this may not always be the case. For example, when the ratio of the edge height to noise is low, $p(\releps|\bar{y}_{1:n})$ is less likely to be Gaussian and in extreme cases may even be multimodal. If the result is bimodal or multimodal then the sample mean and variance may not be a good representation.


\subsection{Covariance Function and Hyperparameter Optimisation} 
\label{sub:hyperparameter_optimisation}
Typically, a covariance function has several hyperparameters, $\theta$. For instance, both the squared-exponential and Mat\'{e}n covariance functions have hyperparameters $\theta = \{\sigma_f,l\}$, where $\sigma_f$ encodes the prior uncertainty and $l$ governs the characteristic smoothness of functions belonging to the GP. 
The covariance function and its hyperparameters can be selected by maximising the marginal log-likelihood \citep{rasmussen2006gaussian}.
For each covariance function considered the hyperparameters are chosen according to
\begin{equation*}
\begin{split}
    \theta_* &= \argmax_\theta \ \log p(y_{1:n} | \theta) \\
    &= \argmax_\theta \ -\frac{1}{2}\left[\log\det(K_y) + Y^{\Transp}K_y^{-1}Y\right]
\end{split}
\end{equation*}
where $Y = \pmat{y_1 & \hdots & y_n}^{\Transp}$, $K_y = K+I\sigma^2$ and
\small
\begin{equation*}\setlength\arraycolsep{1pt} 
\begin{split}
    K = \pmat{A(\lambda_1)k(\lambda_1,\lambda_1)A(\lambda_1)^{\Transp} & \hdots & A(\lambda_1)k(\lambda_1,\lambda_n)A(\lambda_n)^{\Transp} \\
    \vdots & \ddots & \vdots \\
    A(\lambda_n)k(\lambda_n,\lambda_1)A(\lambda_1)^{\Transp} & \hdots & A(\lambda_n)k(\lambda_n,\lambda_n)A(\lambda_n)^{\Transp}}
\end{split}
\end{equation*}
\normalsize
After optimising the hyperparameters for each covariance function, the covariance function that yielded the highest marginal log-likelihood is chosen.

This optimisation can be performed using a gradient-based method, such as the BFGS algorithm presented by \citet{wright1999numerical} with the gradients given by \citet{rasmussen2006gaussian} as
\small
\begin{equation*}
    \frac{\partial \log p(y_{1:n}|\theta)}{\partial\theta_i} = \frac{1}{2}Y^{\Transp}K_y^{-1}\frac{\partial K}{\partial \theta_i}K_y^{-1}Y -\frac{1}{2}\text{tr}\left(K_y^{-1}\frac{\partial K}{\partial \theta_i}\right).
\end{equation*}
\normalsize


\section{Numerical Simulation Study and Error Analysis} 
\label{sec:simulation_demonstration_and_error_analysis}
This section provides a numerical analysis of the accuracy and robustness of the existing approaches and our proposed approach using simulated random trials.
Measurements of the transmission were simulated using Equation~\eqref{eq:santistban_model} with the Bragg-edge shape given by Equation~\eqref{eq:kropff_model} as this provides a good approximation near the region of a Bragg-edge \citep{santisteban2001time}.
These measurements were corrupted with noise according to the noise model determined in Section~\ref{sec:relative_transmission_intensity_noise_analysis}.
To each simulated set of measurements, representing data of a single Bragg-edge, the existing approaches and our proposed approach are applied to estimate the strain and the error between the estimate and the true strain calculated.
In the calculation of strain, the true $\lambda_0$ value was used for the Santisteban and Tremsin methods, and the true stress-free edge profile was used for the cross-correlation method and our approach.

Three different noise levels are investigated; standard deviations given by the analysis with 24 by 24 pixel binning (which will be indicated by $\sigma_{24\times24}$), these standard deviation reduced by a factor of 10 to represent high-quality Bragg-edge data with a high edge height to noise ratio, and these standard deviation increased by a factor of 10 to represent low-quality Bragg-edge data with a low edge height to noise ratio.
For each noise level, \num{10000} random trials were conducted, broken into 100 groups. For each group the parameters of the edge shape were randomly chosen selected from the ranges $\sigma_B=[\num{4.7e-3},\num{1.4e-2}]$, and $\tau=[0,\num{1.3e-2}]$. 
The results are summarised in Table~\ref{tab:simulation_analysis}.

Ideally we would like the methods to produce the following:
\begin{itemize}
    \item A low mean error as this means the results are unbiased.
    \item A low mean magnitude of the errors.
    \item A low standard deviation of the errors, as this would indicate that there is not a large spread of errors.
    \item For the mean of the methods predictions of standard deviation (std) to be close to the calculated standard deviation of the errors, as this indicates the methods can provide an accurate estimate of confidence in the results.
    \item Lastly, for the maximum error to be within a couple of standard deviations of zero, otherwise it indicates the presence of outliers.
\end{itemize}

\newcommand{\ra}[1]{\renewcommand{\arraystretch}{#1}}
\begin{table*}[htb]
\ra{1.3}
\centering
\scalebox{0.8}{
\begin{tabular}{@{}lrrrr@{}}\toprule
& \citet{santisteban2001time} & \citet{tremsin2016investigation} & \citet{ramadhan2018neutron}& Our approach  \\ 
\midrule
Noise level $\sigma_{24\times24}$: \\
\hspace{2.5mm} Error mean           & -53.1123  & \red{-78.16}  & \blue{-32.44}   & 34.05\\
\hspace{2.5mm} Mean magnitude       & \red{244.96}  & 208.50  & \blue{54.02}   & 79.60 \\ 
\hspace{2.5mm} Maximum              & 2706.73 & \red{2846.05} & 1216.81  & \blue{421.18}\\
\hspace{2.5mm} Standard deviation   & 424.55  & 354.55  & \blue{62.70}   & 94.87\\
\hspace{2.5mm} Mean predicted std   & 175.92  & 220.90  & \red{12.76}   & \blue{88.89}\\
Noise level $\sfrac{\sigma_{24\times24}}{10}$: \\
\hspace{2.5mm} Error mean           & \blue{14.25}  & \red{-21.15} & -17.82 & 15.19 \\
\hspace{2.5mm} Mean magnitude       & 27.45  & \red{57.01}  & \blue{25.87}  & 26.07 \\ 
\hspace{2.5mm} Maximum              & \red{2359.47} & 454.78 & \blue{83.44} & 94.26 \\
\hspace{2.5mm} Standard deviation   & \red{111.91} & 77.91  & \blue{25.46}  & 28.77  \\
\hspace{2.5mm} Mean predicted std   & \red{16.69}  & 24.48  & \blue{11.01}  & 10.89 \\
Noise level $10\sigma_{24\times24}$: \\ 
\hspace{2.5mm} Error mean           & \red{-559.56} & -462.14 & \blue{-32.11}  & -34.19  \\
\hspace{2.5mm} Mean magnitude       & \red{1192.65} & 1102.45 & \blue{435.53}  & 564.61  \\ 
\hspace{2.5mm} Maximum              & 6985.37 & \red{8627.56} & 5762.89 & \blue{4468.49}\\
\hspace{2.5mm} Standard deviation   & \red{1515.31} & 1497.60 & \blue{570.36} & 722.63 \\
\hspace{2.5mm} Mean predicted std   & 2510.91 & \red{\num{4.15e7}} & 125.48 & \blue{946.17} \\
\bottomrule
\end{tabular}}
\caption{Analysis of the errors, in $\mu$-strain, given by the existing approaches and our proposed approach from the simulated random trials. The best value of each metric for each noise level is coloured \blue{blue} and the worst \red{red}. It should be noted that for the cross-correlation method it was necessary to manually tune the smoothing parameters for each noise level to achieve the best results.}
\label{tab:simulation_analysis}
\end{table*}

The results presented in Table~\ref{tab:simulation_analysis} indicate both the cross-correlation method and our proposed approach achieve good mean absolute error, mean error, and standard deviation of error. 
Significantly, applying these methods to the noisiest data ($10\sigma_{24\times24}$) sets yields results almost as accurate as the Santisteban and Tremsin methods applied to the data sets with the standard noisy level ($\sigma_{24\times24}$).
For the cross-correlation procedure this required manually tuning the smoothing parameters for each noise level --- this could be challenging to do well for a large set of strain images with varying Bragg-edge height to noise ratios.

When well-tuned, the cross-correlation method gives slightly lower mean absolute error, mean error, and standard deviation of error than our proposed approach. However, our proposed approach yields a significantly more accurate prediction of standard deviation, and a lower maximum error on all but the lowest noise simulations. This higher maximum error for the cross-correlation method is indicative that the method occasionally produces outlying estimates of strain.

It is our hypothesis that the outliers are due to the need to perform a numerical derivative. A significant challenge when applying this approach is determining the fit window and polynomial order that produce the lowest bias, mean absolute errors, and maximum error. In general, it was found that increasing the fit window decreased the standard deviation of the errors and reduced the likelihood of outliers, but after a certain point would increase the bias. Since the results here are from a simulated numerical study, the authors had the true values of strain to aid in tuning these parameters; it is unclear what the best way to tune these parameters is in general.
A secondary challenge is choosing a good starting guess for the parameters when fitting the Voigt function, since a bad starting guess could result in the optimisation procedure finding a local rather than global minima.

Additionally, the standard deviations predicted by the cross-correlation method are, for the most part, far too low. This is likely due to the fact that the method has three stages (derivatives are taken, cross-correlation is performed, a parametric function is fit) whereas the prediction of standard deviation is based only on the final fitting stage. As a consequence, the user may be given the impression that the results are accurate even if this is not the case. For example, the average predicted standard deviation is only marginally higher for the standard noise simulations ($12.76\mu\text{strain}$) than for the low noise simulations ($11.01\mu\text{strain}$), whereas the mean error is twice as large and the maximum error is far greater. In contrast, our proposed approach gives predictions of the standard deviation that are relatively accurate for the standard noise and high noise simulations.

\begin{figure}[htb]
\centering
\subcaptionbox{\citet{santisteban2001time}}
{\includegraphics[width=0.49\linewidth]{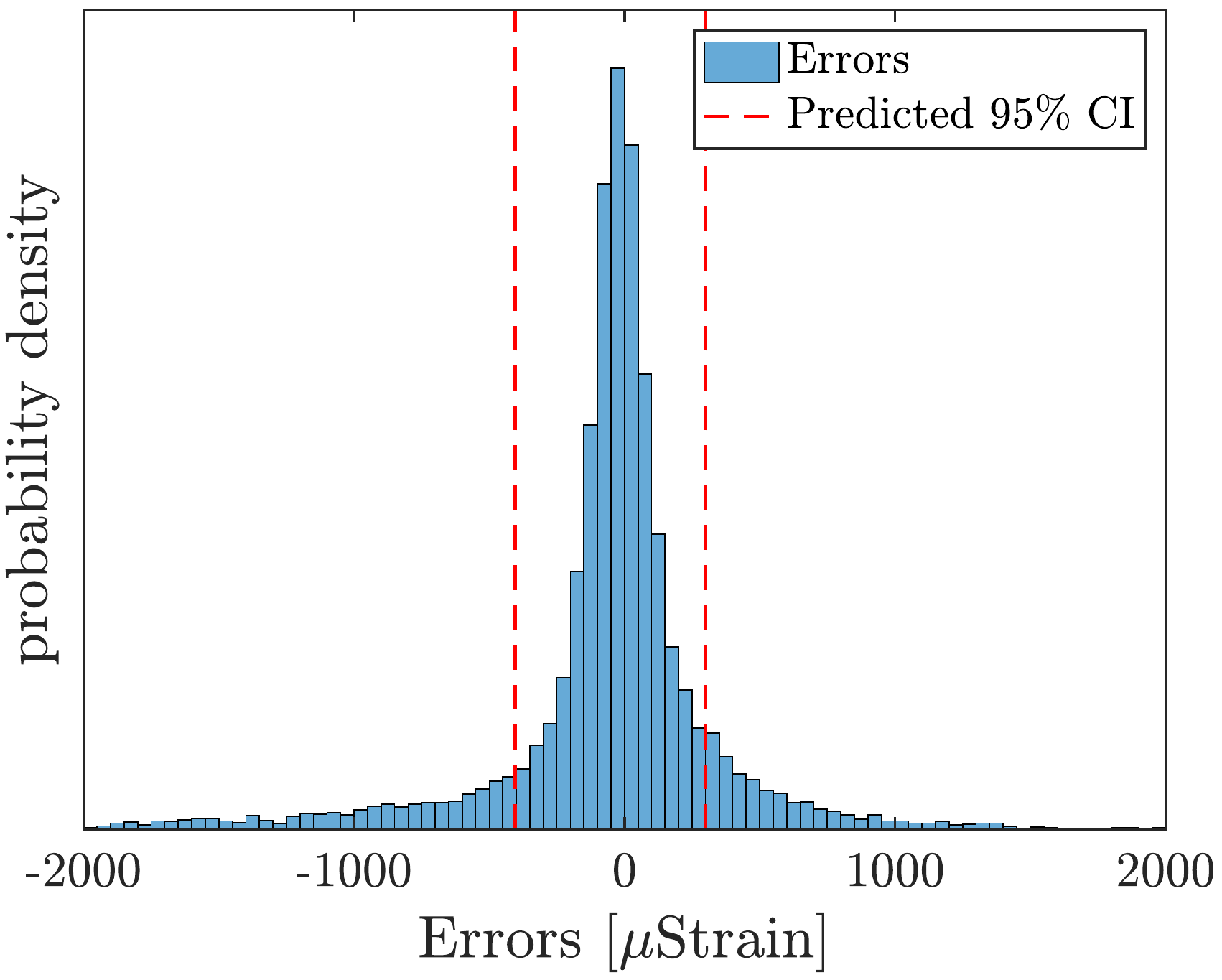}}
\subcaptionbox{\citet{ramadhan2018neutron}}
{\includegraphics[width=0.49\linewidth]{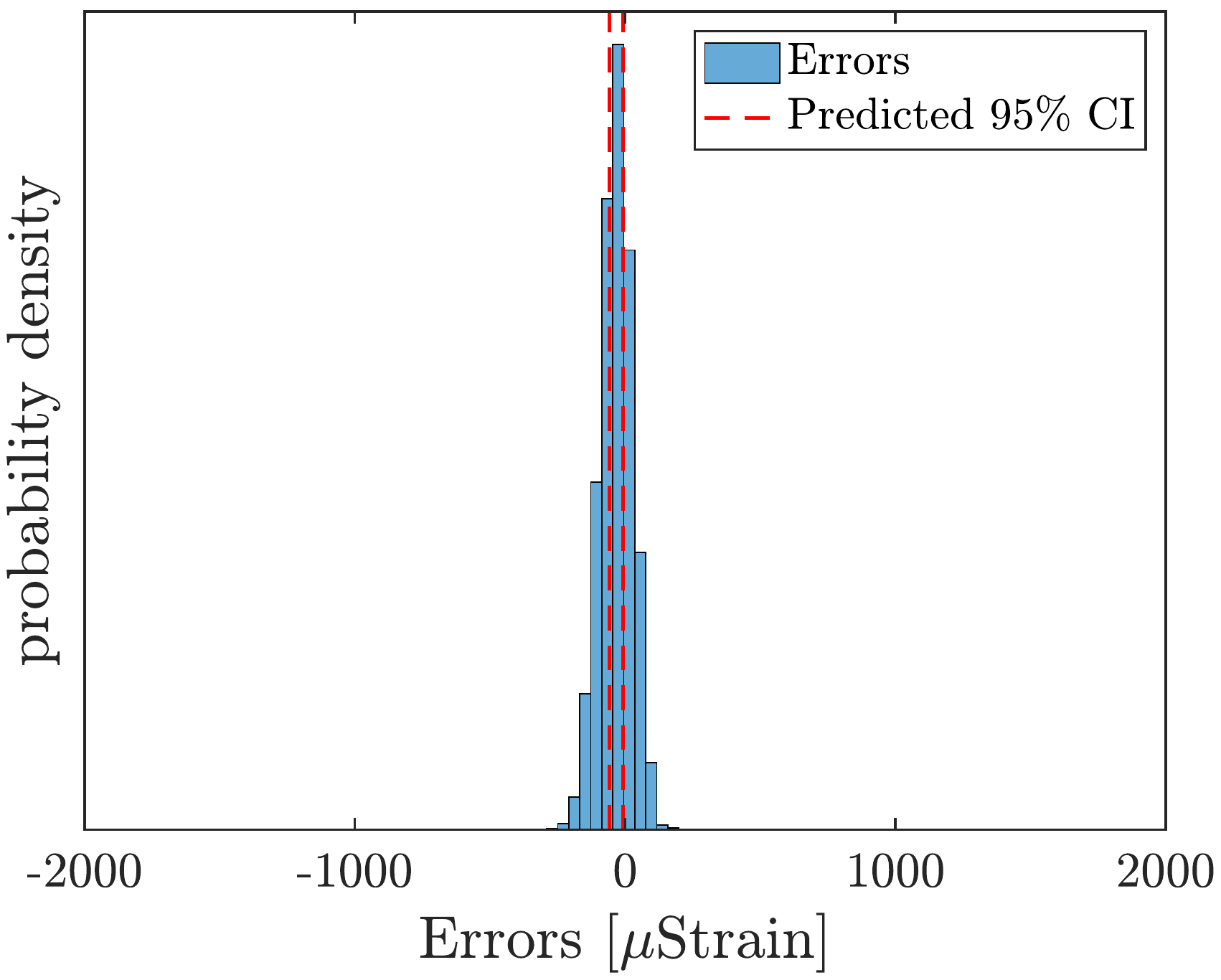}} \\
\subcaptionbox{\citet{tremsin2016investigation}}
{\includegraphics[width=0.49\linewidth]{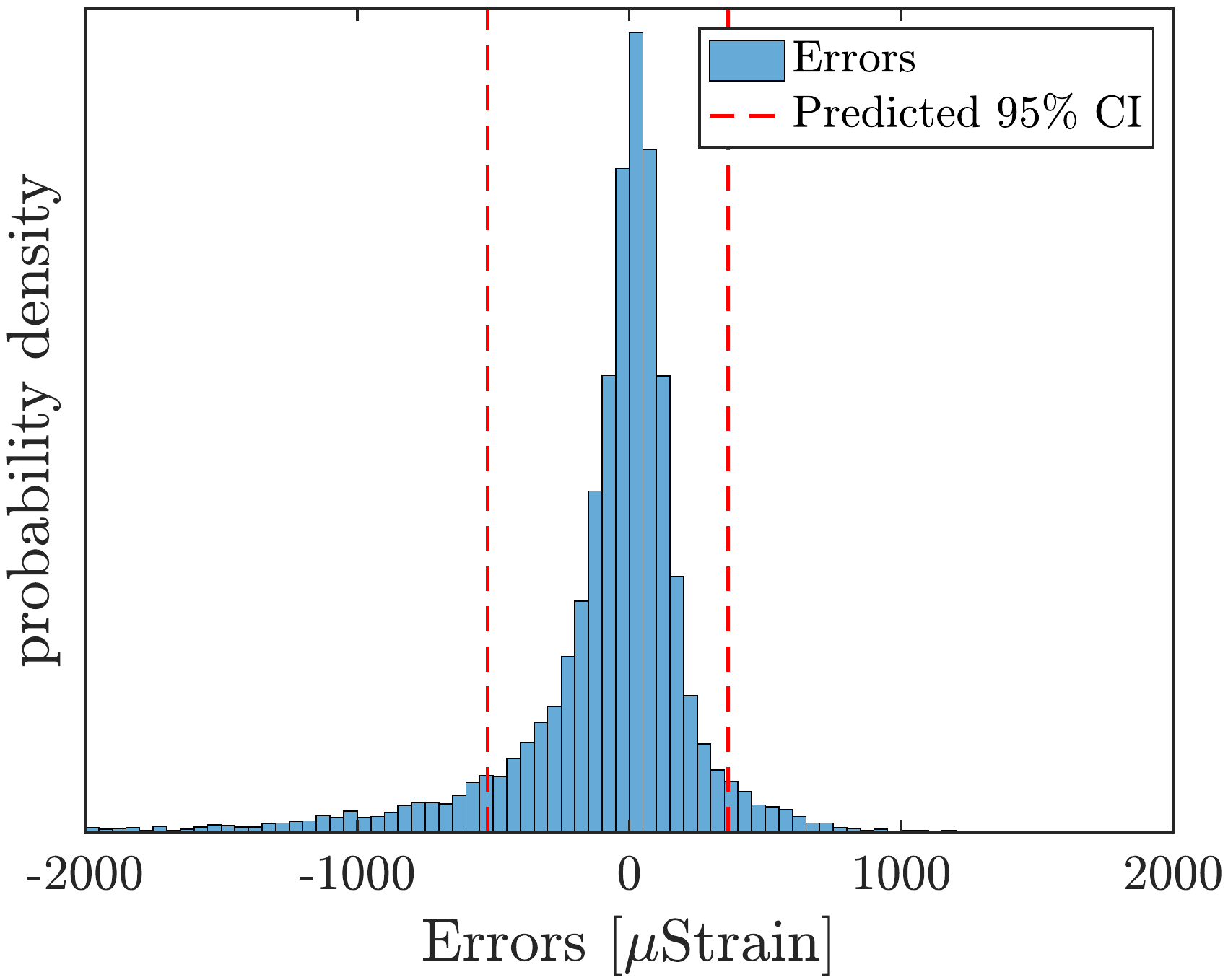}}
\subcaptionbox{Our approach}
{\includegraphics[width=0.49\linewidth]{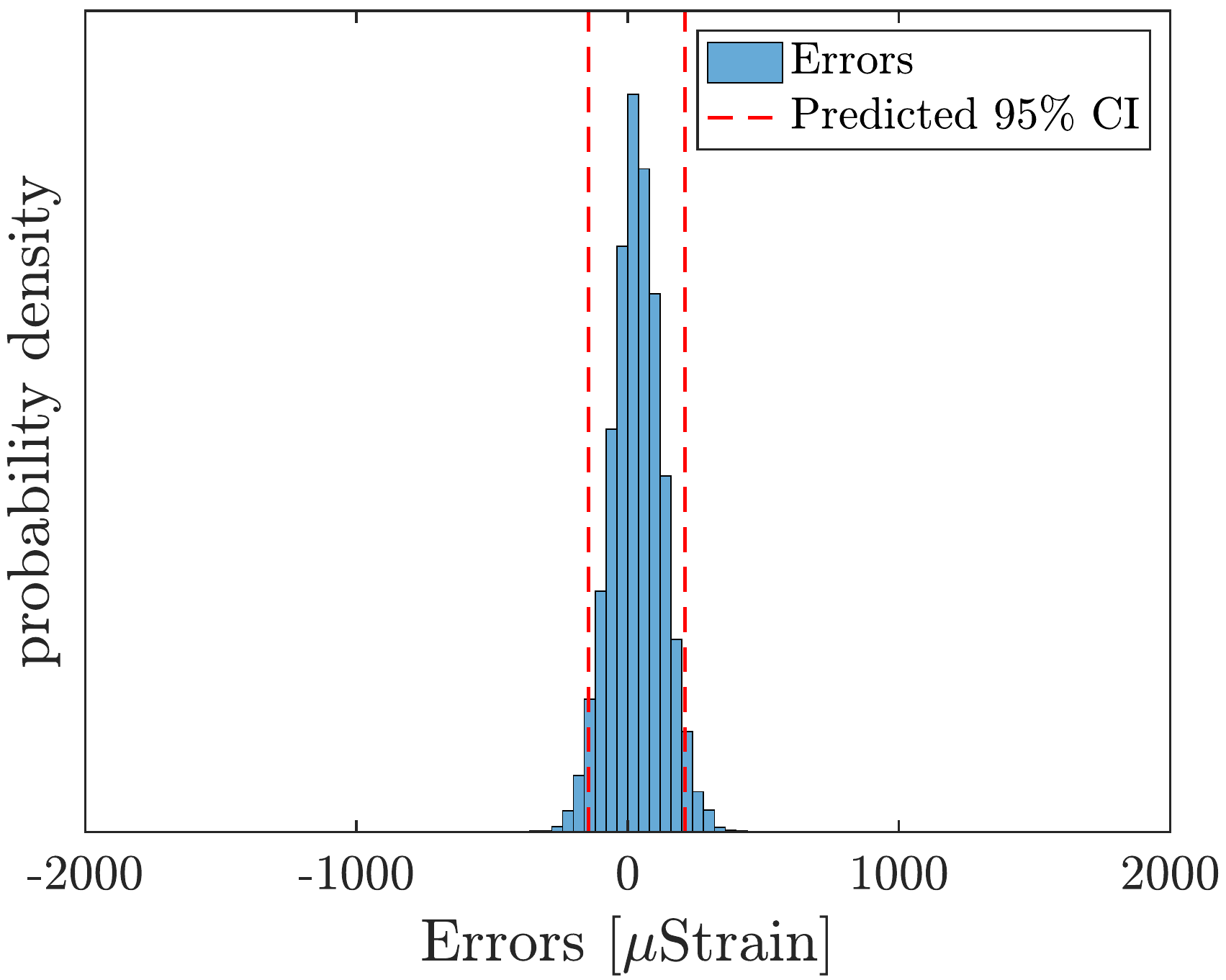}}
\caption{Error distributions from the random trial using the noise model given in Section~\ref{sec:relative_transmission_intensity_noise_analysis} with averaging over 24 by 24 pixels. The distributions are shown for the existing approaches and the proposed approach along with each methods mean predicted confidence interval. }
\label{fig:error_dists}
\end{figure}

The error distributions calculated for the $\sigma_{24\times24}$ noise level are shown in Figure~\ref{fig:error_dists} along with the mean predicted confidence interval given by each method. Confidence intervals have been calculated from the predicted standard deviation using a Gaussian assumption, whereby $\mu\pm2\sigma$ gives the $95\%$ region.
These histograms show that the error distributions from the Santisteban, and Tremsin have heavy tails.
The histogram does not show the outliers, maximum error magnitudes, of the cross-correlation method as they do not occur at high enough density to be clearly displayed..
Further, a significant number of the errors for the Santisteban and Tremsin methods are outside the average predicted $95\%$ confidence intervals, while for the cross-correlation method the prediction of this region is on average far too small --- to the extent that zero-error is right on the edge of the $95\%$ confidence interval. In contrast, our proposed approach has a roughly Gaussian distribution for the errors with a reasonably accurate $95\%$ confidence interval.

\section{Experimental Data Demonstration} 
\label{sec:experimental_data_demonstration}
The existing methods and our proposed approach are demonstrated on two sets of experimentally collected neutron transmission Bragg-edge data. The first set is of a AlSiCP metal matrix composite plate and was previously used to demonstrate the cross-correlation method. For this data set, there are corresponding diffraction strain measurements allowing for a quantitative comparison to be made.
The second data set is part of a strain tomography data set and qualitatively compares the ability of the methods to produce strain images that can be used for strain tomography.

\subsection{Data from Sample with Significant Texture} 
\label{sub:data_from_cross_correlation_paper}
The proposed approach is demonstrated on experimental neutron transmission Bragg-edge data described in detail by \citet{ramadhan2018neutron} and compared to results from the existing approaches and diffraction strain measurements \citet{ramadhan2019characterization}.
This data is taken of an AlSiCP metal matrix composite (MMC) plate composed of an AL 2124 matrix and pure silicon with dimensions of $\SI{15}{\milli\meter}$ in $z$ and $\SI{35}{\milli\meter}$ in $x$ and $y$. 
This data is an appropriate experimental test set as it was previously used to compare the existing approaches for strain estimation.

Data was collected using a microchannel plate (MCP) detector \citep{tremsin11} which has a 512 by 512 array of $\SI{55}{\micro\meter}$ by $\SI{55}{\micro\meter}$ pixels, a five hour measurement time, and the samples $x$ dimension aligned with the neutron beam.
The data was then averaged over regions of $\SI{1}{\milli\meter}$ in the $z$-direction and $\SI{20}{\milli\meter}$ in the $y$-direction, for which there is minimal strain variation.
After averaging, the Bragg-edge data has minimal noise.

\begin{figure}[tb]
    \centering
    \includegraphics[width=1.0\linewidth]{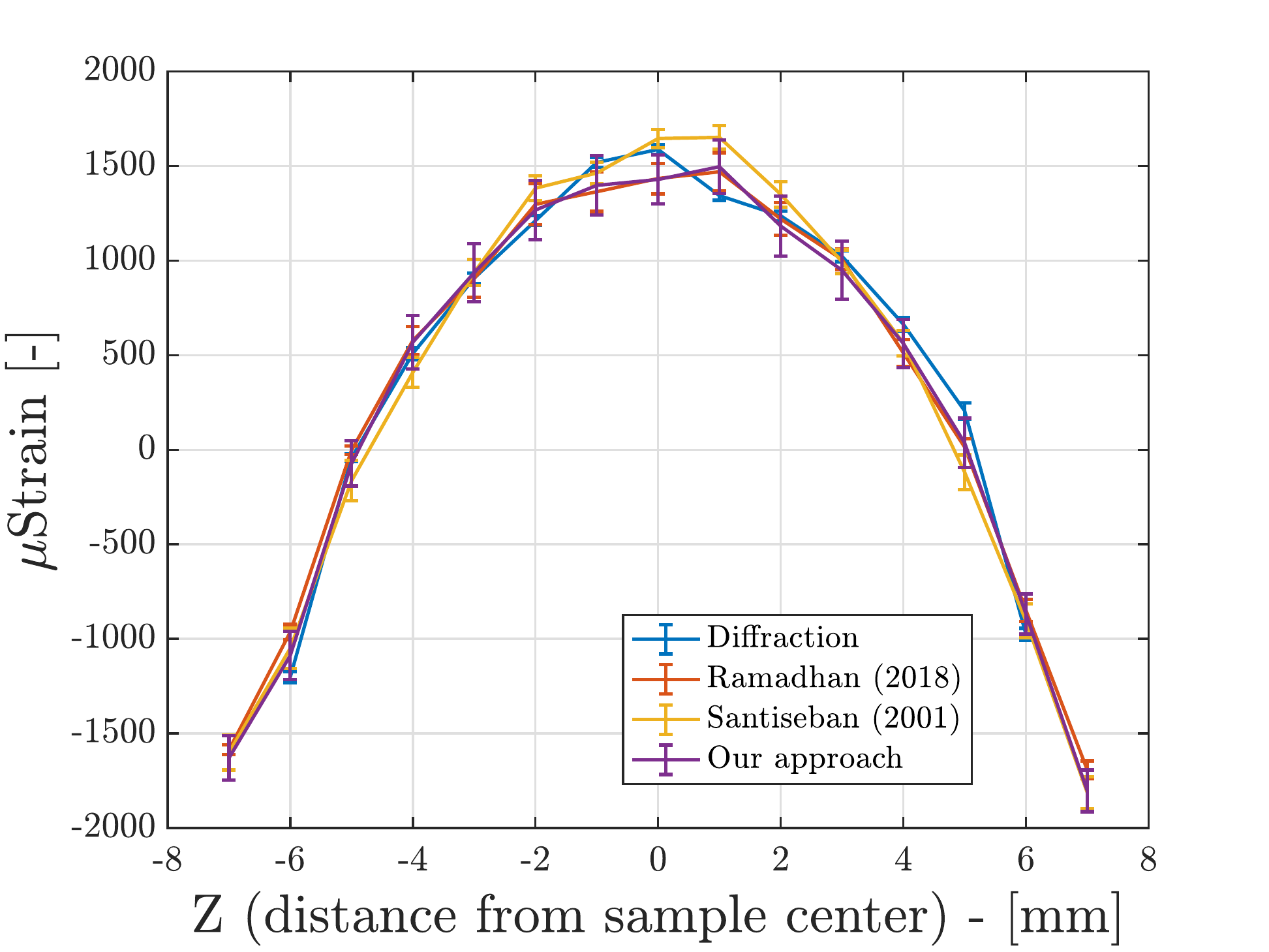}
    \caption{Strain estimates from Bragg-edge neutron transmission data of a AlSiCP MMC plate. The estimated strain values and the one standard deviation confidence intervals are shown. Also shown for comparison are results from neutron diffraction \citep{ramadhan2019characterization}.}
    \label{fig:cross_corr_data_comp}
\end{figure}

The strain is estimated by applying the methods to the Aluminium $(111)$ Bragg-edge and the results are shown in Figure~\ref{fig:cross_corr_data_comp}; the estimated strains and the one standard deviation error bars are shown.
As in \citet{ramadhan2018neutron}, it was found that applying the approach by \citet{tremsin2016investigation} yielded very similar but slightly worse results than the Santisteban function and so they are not shown.

Calculating the mean absolute difference between the diffraction measurements and the estimated strains from transmission data gives; $94.42\pm{17.28}\,\mu\text{Strain}$ for the cross-correlation method, $105.63\pm30.77\,\mu\text{Strain}$ for our approach, and $145.73\pm19.2216\,\mu\text{Strain}$ for the approach by \citet{santisteban2001time}.
The one standard deviation range on the mean difference has contributions from both the predicted standard deviation of estimated strain from neutron transmission and diffraction, as neutron diffraction are also a measurement and not ground truth.
Given that both the cross-correlation and our proposed approach have a mean absolute difference one standard deviation smaller than the Santisteban function, we can tentatively conclude that they perform better on this data. With the cross correlation performing marginally better.

This is expected as the Al 2124 metal matrix is reported to have significant preferred crystal orientation (texture), which distorts the Al $(111)$ edge shape \citep{ramadhan2018neutron}.
As a result, the parametric edge shapes used by \citet{santisteban2001time} and \citet{tremsin2016investigation} are not a good model for the edge shape. Hence, the use of the cross-correlation method and our proposed non-parametric approach are beneficial. 


\subsection{Strain Imaging Example} 
\label{sub:strain_imaging_example}
\begin{figure}[htb]
    \centering
    \includegraphics[trim={150 0 150 0}, clip = true, width = 0.6\linewidth]{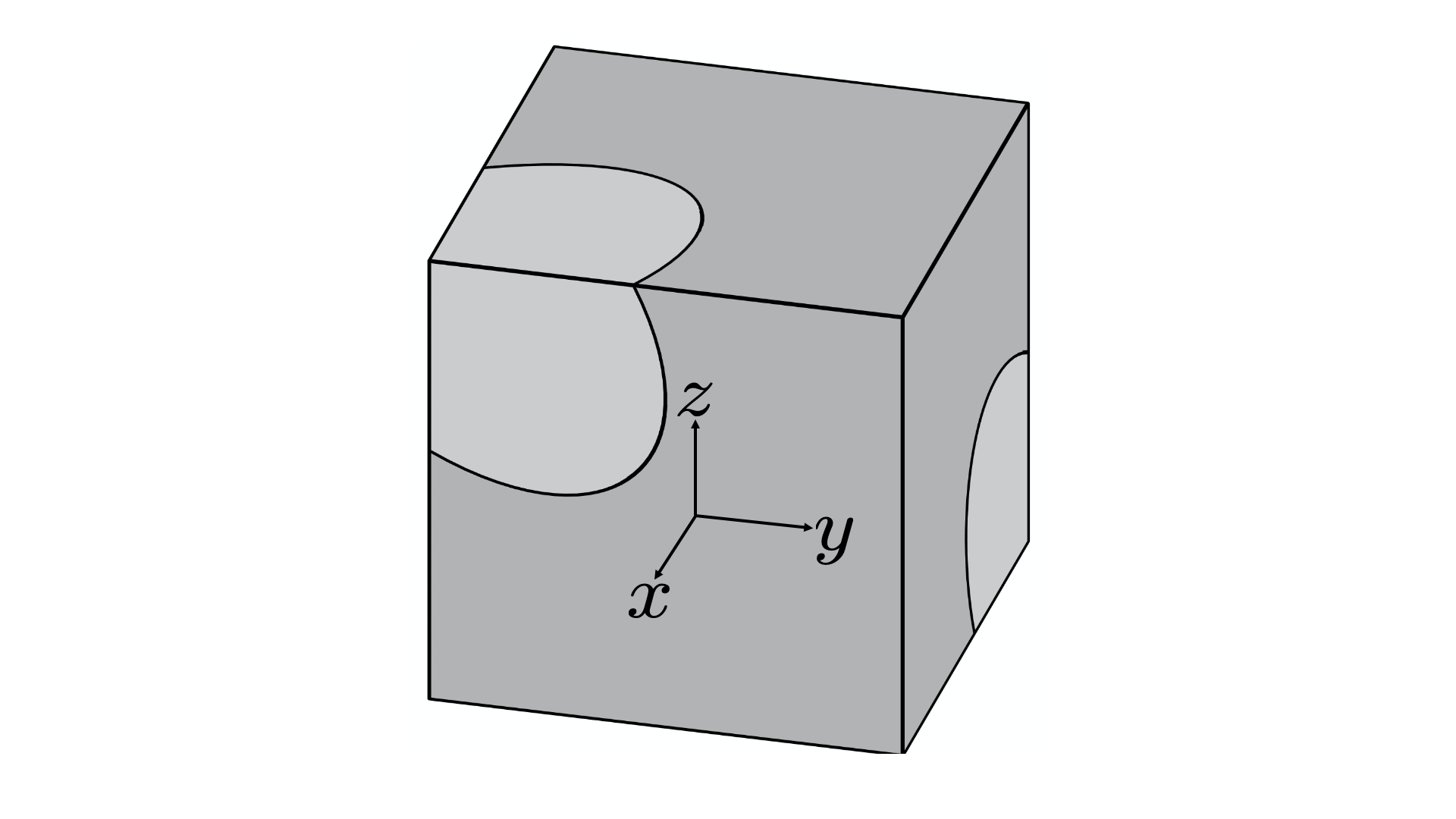}
    \caption{The cube (dark grey) and plug (light grey) sample assembly. Figure taken from \citep{hendriks2019tomographic}.}
    \label{fig:samle}
\end{figure}

The proposed approach is demonstrated on experimental neutron transmission Bragg-edge data collected on the RADEN energy-resolved-neutron-imaging instrument at J-PARC \citep{shinohara2015commissioning,shinohara2016final}, and compared with existing methods. The sample was manufactured primarily for triaxial strain tomography, it is EN26 steel (medium carbon, low alloy) consisting of a $17 \times 17 \times 17~\si{\milli\metre}$ steel cube with a precision ground hole of diameter 12~\si{\milli\metre} along the diagonal into which an EN26 steel plug was fit with an interference of $40\pm2$ \si{\micro\metre} interference fit, i.e., shrink fit --- shown in Figure~\ref{fig:samle}. 
This sample was manufactured to provide a reference for three dimensional strain tomography.
Strain images were measured using a micro-channel plate detector ($512~\times~512$ pixels, 55 \si{\micro\metre} per pixel) \citep{tremsin11} at a distance of 17.9 \si{\metre} from the source. At the time of the experiment (February 2020) the source power was 500 \si{\kilo\watt}.



The data-set was collected for tomographic reconstruction of a strain field, for which many projections are  collected in a limited amount of beam-time and as little averaging over pixels as possible is desired to obtain a high resolution strain reconstruction. Both of these factors contribute to noise levels in the data, motivates Bragg-edge fitting methods which perform well in the presence of noise and provide an accurate measure of certainty which can be used by strain tomography methods to weight the importance of different strain measurements.

\begin{figure}[h]
    \centering
    \begin{minipage}[b]{0.49\linewidth}
        \includegraphics[width = 0.9\textwidth]{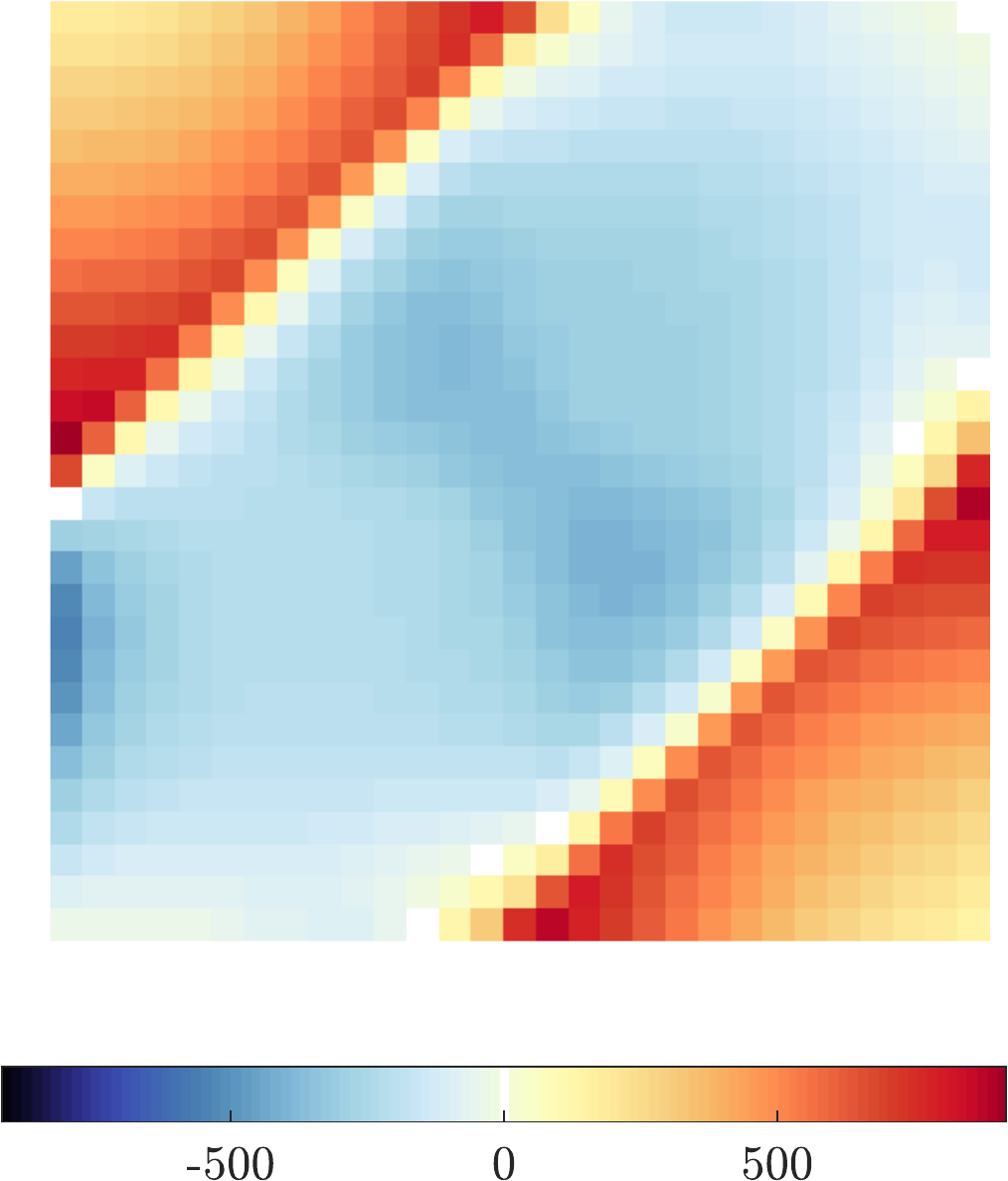}
        \subcaption{~}
    \end{minipage}
        \begin{minipage}[b]{0.49\linewidth}
        \includegraphics[width = \textwidth]{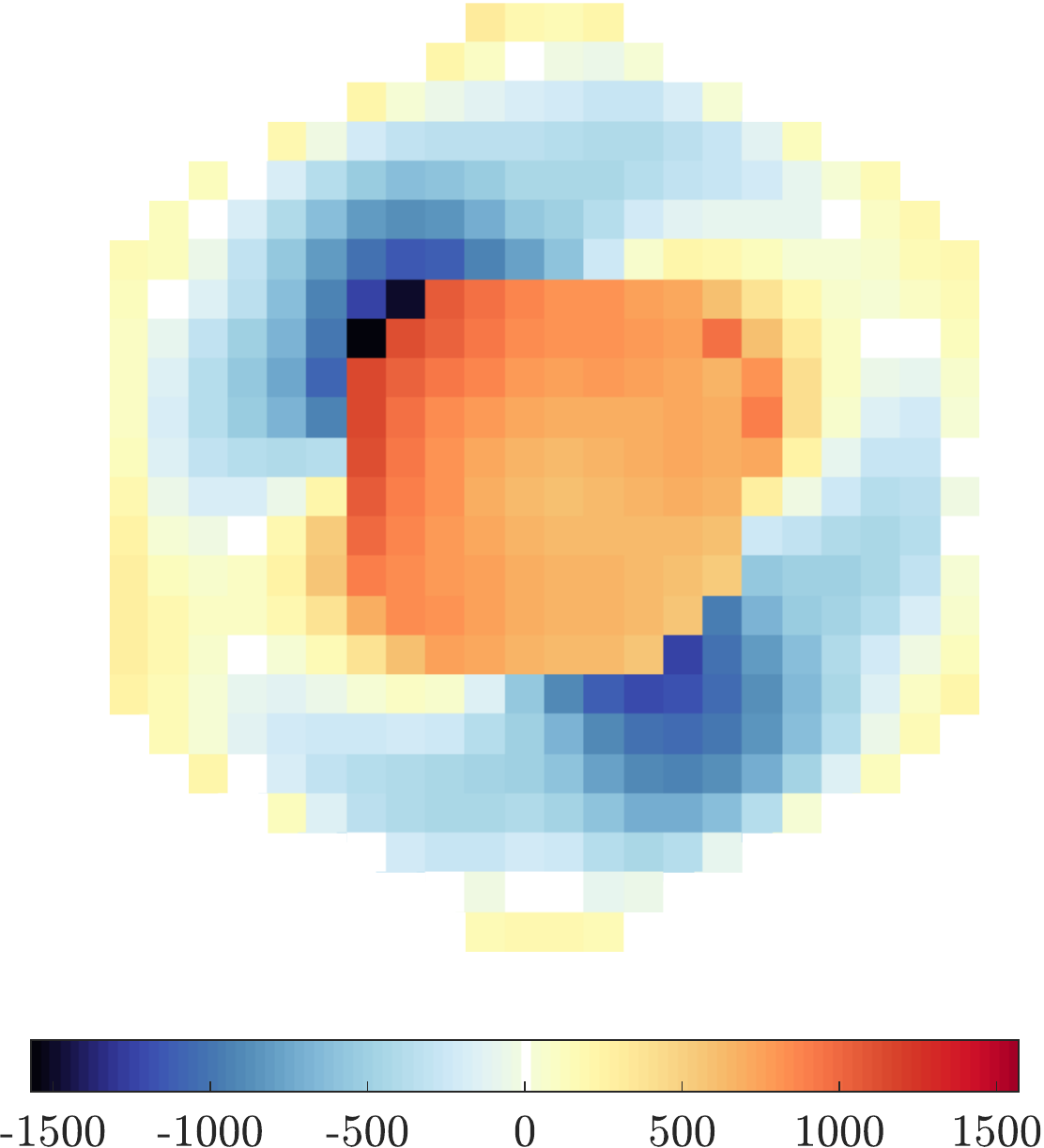}
    \subcaption{~}
    \end{minipage}
    \caption{Simulated strain images. (a) shows a projection aligned with the face of the cube, referred to as projection 1, it has been produced at resolution which emulates averaging over 10$\times$10 pixels. (b) shows a projection aligned with the axis of the plug, referred to as projection 2, it has been produced at a resolution which emulates averaging over 18$\times$18 pixels.}
    \label{fig:FEA}
\end{figure}

Figure~\ref{fig:FEA} shows two simulated strain images generated using finite element results. The first projection is aligned with the face of the cube and the second is aligned with the axis of the plug. Lacking secondary experimental data for this sample, e.g., diffraction, the FEA results will provide a point of reference for each method's performance. 

\begin{figure*}[h!]
    \centering
    \begin{tabular}{L{0.02\textwidth} L{0.2\textwidth} L{0.01\textwidth} L{0.18\textwidth} L{0.01\textwidth} L{0.19\textwidth} L{0.01\textwidth} L{0.2\textwidth} L{0.01\textwidth} }
    & \citet{santisteban2001time} &&\citet{tremsin2016investigation}&&\citet{ramadhan2018neutron}&&Proposed method&\\
    \includegraphics[width = \textwidth]{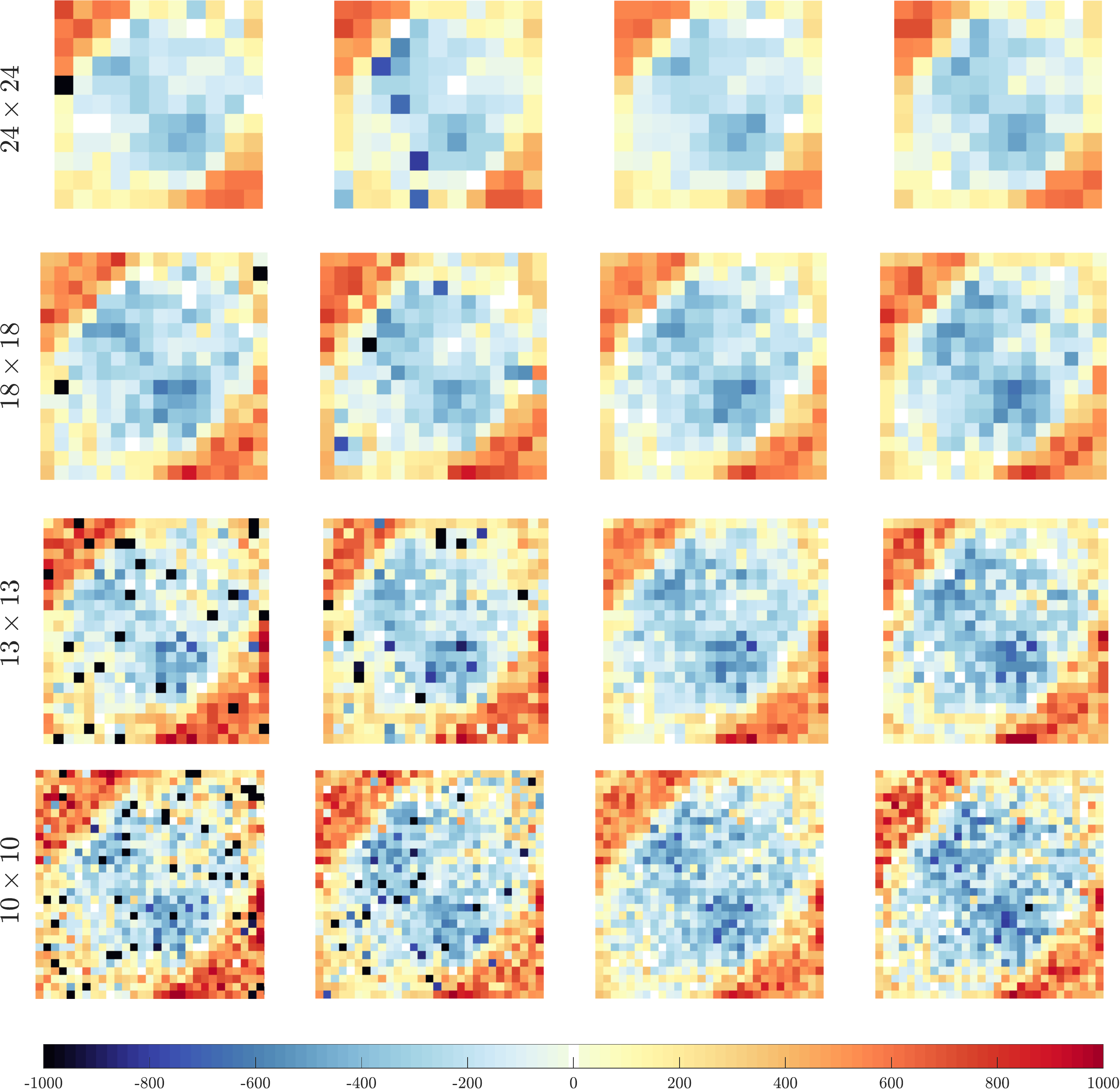}
    \end{tabular}
    \caption{Comparison of strain images using a single projection. The strain images shown were produced using each method for 3 different pixel groupings. All units are $\mu$-strain.}
    \label{fig:StrainImageGrid}
\end{figure*}

Strain is estimated by applying the methods to the (110) Bragg-edge and results are shown in Figure~\ref{fig:StrainImageGrid}.
It is apparent that both the cross-correlation method and our proposed approach perform substantially better in the presence of higher levels of noise than the other methods. Recalling how pixel grouping affects measurement noise from Section~\ref{sec:relative_transmission_intensity_noise_analysis}, this can be observed in the degree of agreement between each strain image produced by the same method. Both our approach and the cross-correlation method experience good agreement between the 24-by-24, 18-by-18 and 13-by-13 pixel groupings. While both of the other methods experience significantly many more outliers, indicated by the black and dark red pixels which saturate the colour scale, with the introduction of higher noise levels. 

We can also assess the quality of each image by observing its \emph{smoothness}, giving a rough indication of how noisy each image is. Large unexpected jumps in strain between neighbouring pixels give an indication of a \emph{noisy} image and smooth strain images give an indication that the image is \emph{less noisy}. These unexpected jumps in strain are even present in the strain images produced by the Santisteban and Tremsin method using the relatively noise free 24-by-24 pixel grouping, where-as for both our proposed approach and the cross correlation method this \emph{noise} is not present in the image until significantly more noise is present in the data, e.g., the 13-by-13 pixel grouping.

Table~\ref{tab:AverageCI} shows the mean predicted standard deviation for each of the Bragg-edges fitted to produce Figure~\ref{fig:StrainImageGrid}. As each ray has the same irradiated length in the shown projection, and therefore very similar noise levels, the average confidence interval should give a good indication of each methods confidence interval for the whole image. These results reflect the ones from Section~\ref{sec:simulation_demonstration_and_error_analysis}, the cross-correlation method is consistently overconfident in its results, where-as our method's predicted confidence interval reflects the addition of noise introduced by averaging across fewer pixels.

\begin{table}[htb]
\ra{1.3}
\centering
\caption{Mean predicted standard deviation, in $\mu$-strain, given by each method for the strain images shown in Figure~\ref{fig:StrainImageGrid}. } 
\label{tab:AverageCI}
\scalebox{0.78}{
\begin{tabular}{R{0.21\linewidth} R{0.23\linewidth} R{0.19\linewidth} R{0.21\linewidth} R{0.18\linewidth}}\toprule
Resolution& \citet{santisteban2001time} & \citet{tremsin2016investigation} & \citet{ramadhan2018neutron}& Our approach  \\ 
\midrule
$24\times24$& 133.69& 64.71& 5.77& 83.81\\
$18\times18$& 174.14& 78.69& 6.89& 110.45\\
$13\times13$& 231.96& 100.85& 8.61& 150.20\\
$10\times10$& 290.75& 115.54& 10.77& 190.35\\
\bottomrule
\end{tabular}}
\end{table}

\begin{figure}[h]
    \centering
        \begin{tabular}{L{0.01\linewidth} L{0.42\linewidth}  L{0.42\linewidth} L{0.02\linewidth} }
    &\citet{ramadhan2018neutron} & Proposed method& \\
    \multicolumn{4}{l}{
    \includegraphics[trim = {0 0 0 8mm},clip=true,width = 0.95\linewidth]{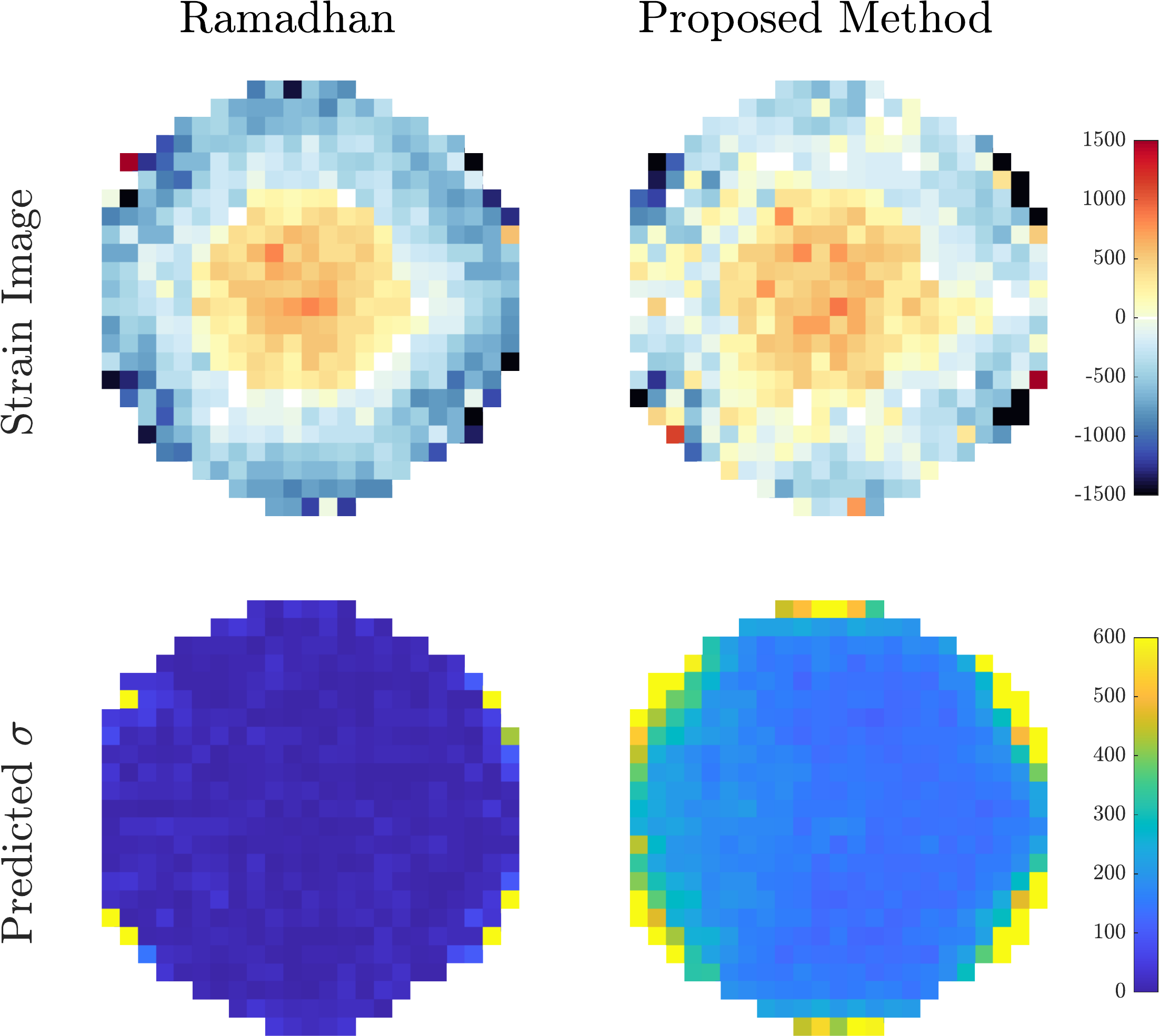}
    }
    \end{tabular}
    \caption{Strain images of projection 2 produced by both the cross-correlation method and our proposed approach. All units are  $\mu$-strain.}
    \label{fig:projection2}
\end{figure}

A second strain imaging projection shown in Figure~\ref{fig:projection2} depicts the results of both the cross-correlation method and our proposed approach. In this projection the beam passes through varying amounts of material, ranging from 29~\si{\milli \metre} in the centre of the image to zero at the perimeter, for this reason the signal to noise ratio is better in the centre and worse near the edges, this should be reflected in the standard deviation predicted by each method. Although the strain image produced by the cross correlation method appears to be slightly less noisy the associated predicted standard deviation is over confident, and has virtually no variation across the image. In contrast, the standard deviation predicted by our proposed approach increases from the centre of the image to the edge. The ability to accurately estimate the confidence interval for Bragg-edge fitting is imperative for some strain tomography methods \citep{hendriks2019tomographic}.

\section{Conclusion and Future Work} 
\label{sec:conclusion}
In this paper we have focussed on the problem of estimating strain from energy resolved neutron transmission Bragg-edge data. Methods for estimating strain from this data can be used to generate strain images allowing the residual stress and strain inside a sample to be studied. They can also be used as an intermediate step in strain tomography. For both these applications, the methods used should accurately estimate the strain and give a reliable measure of certainty even when the data is noisy.

A novel Bayesian non-parametric approach is proposed for estimating strain from this data. As part of this approach, the standard deviation of the strain can be accurately predicted using a Monte Carlo method.
Additionally, the non-parametric nature of this approach is beneficial when the Bragg-edge shape is distorted, such as when the the sample has significant preferred crystal orientation.

A numerical simulation study and two experimental data sets were used to compare the existing approaches and our proposed approach. The results show that both our proposed approach and the cross-correlation method provide more accurate results than the Tremsin and Santisteban method when the data is noisy. The numerical simulation study indicates that the cross-correlation approach gives, on average, slightly lower error magnitudes than our proposed approach. However, this is at the cost of larger maximum errors and inaccurate predictions of standard deviation.

The main challenge in applying the cross correlation method is the need to manually tune the smoothing for the numerical derivatives. This numerical derivative is possibly the cause of any outliers produced by this method. 

A path forward would be to combine parts of our proposed approach and the cross-correlation method. Our proposed approach provides a method for computing exact, rather than numerical, derivatives and through hyperparameter optimisation removes the need for manual tuning. Once the derivatives are calculated using our approach, the remainder of the cross-correlation method could be applied. Additionally, this would potentially allow accurate prediction of the standard deviation by using Monte Carlo sampling of the derivatives and pushing these samples through the rest of the procedure.


\section*{Acknowledgements} 
\label{sec:acknowledgements}
This work was supported by the Australian Research Council through a Discovery Project Grant No. DP170102324. Access to the RADEN instrument was made possible through the user-access program of J-PARC (J-PARC Long Term Proposal No. 2017L0101).

\bibliography{References}

\begin{thebibliography}{26}
\providecommand{\natexlab}[1]{#1}
\providecommand{\url}[1]{\texttt{#1}}
\expandafter\ifx\csname urlstyle\endcsname\relax
  \providecommand{\doi}[1]{doi: #1}\else
  \providecommand{\doi}{doi: \begingroup \urlstyle{rm}\Url}\fi

\bibitem[Abramowitz and Stegun(1964)]{abramowitz1964handbook}
M.~Abramowitz and I.~A. Stegun.
\newblock \emph{Handbook of mathematical functions. 1965}, pages 84--85.
\newblock Dover Publications, New York, 1964.

\bibitem[Bragg and Bragg(1913)]{bragg1913reflection}
W.~H. Bragg and W.~L. Bragg.
\newblock The reflection of {X}-rays by crystals.
\newblock \emph{Proceedings of the Royal Society of London. Series A,
  Containing Papers of a Mathematical and Physical Character}, 88\penalty0
  (605):\penalty0 428--438, 1913.

\bibitem[Geyer(2007)]{geyer2007fisher}
C.~J. Geyer.
\newblock Stat 5102 {N}otes: {F}isher {I}nformation and {C}onfidence
  {I}ntervals {U}sing {M}aximum {L}ikelihood.
\newblock Lecture notes, 2007.
\newblock URL \url{https://www.stat.umn.edu/geyer/s06/5102/notes/fish.pdf}.

\bibitem[Gorry(1990)]{gorry1990general}
P.~A. Gorry.
\newblock General least-squares smoothing and differentiation by the
  convolution (savitzky-golay) method.
\newblock \emph{Analytical Chemistry}, 62\penalty0 (6):\penalty0 570--573,
  1990.

\bibitem[Gregg et~al.(2018)Gregg, Hendriks, Wensrich, Wills, Tremsin, Luzin,
  Shinohara, Kirstein, Meylan, and Kisi]{gregg2018tomographic}
A.~W.~T. Gregg, J.~N. Hendriks, C.~M. Wensrich, A.~G. Wills, A.~S. Tremsin,
  V.~Luzin, T.~Shinohara, O.~Kirstein, M.~H. Meylan, and E.~H. Kisi.
\newblock Tomographic {R}econstruction of {T}wo-{D}imensional {R}esidual
  {S}train {F}ields from {B}ragg-{E}dge {N}eutron {I}maging.
\newblock \emph{Physical Review Applied}, 10\penalty0 (6):\penalty0 064034,
  2018.
\newblock \doi{10.1103/PhysRevApplied.10.064034}.

\bibitem[Hendriks(2020)]{hendriksThesis2020}
J.~N. Hendriks.
\newblock \emph{Probabilistic {M}odelling and {E}stimation of {E}lastic
  {S}train from {D}iffraction-{B}ased {M}easurements}.
\newblock PhD thesis, University of Newcastle, 2020.
\newblock URL \url{http://hdl.handle.net/1959.13/1411479}.

\bibitem[Hendriks et~al.(2019)Hendriks, Gregg, Jackson, Wensrich, Wills,
  Tremsin, Shinohara, Luzin, and Kirstein]{hendriks2019tomographic}
J.~N. Hendriks, A.~W.~T. Gregg, R.~R. Jackson, C.~M. Wensrich, A.~Wills, A.~S.
  Tremsin, T.~Shinohara, V.~Luzin, and O.~Kirstein.
\newblock Tomographic reconstruction of triaxial strain fields from
  {B}ragg-edge neutron imaging.
\newblock \emph{Phys. Rev. Materials}, 3:\penalty0 113803, Nov 2019.
\newblock \doi{10.1103/PhysRevMaterials.3.113803}.

\bibitem[Jidling et~al.(2018)Jidling, Hendriks, Wahlstr{\"o}m, Gregg,
  Sch{\"o}n, Wensrich, and Wills]{jidling2018probabilistic}
C.~Jidling, J.~N. Hendriks, N.~Wahlstr{\"o}m, A.~Gregg, T.~B. Sch{\"o}n,
  C.~Wensrich, and A.~Wills.
\newblock Probabilistic modelling and reconstruction of strain.
\newblock \emph{Nuclear Instruments and Methods in Physics Research Section B:
  Beam Interactions with Materials and Atoms}, 436:\penalty0 141 -- 155, 2018.
\newblock \doi{10.1016/j.nimb.2018.08.051}.

\bibitem[Kropff et~al.(1982)Kropff, Granada, and Mayer]{kropff1982bragg}
F.~Kropff, J.~R. Granada, and R.~E. Mayer.
\newblock The {B}ragg lineshapes in time-of-flight neutron powder spectroscopy.
\newblock \emph{Nuclear Instruments and Methods in Physics Research},
  198\penalty0 (2-3):\penalty0 515--521, 1982.

\bibitem[Mostafavi et~al.(2017)Mostafavi, Collins, Peel, Reinhard, Barhli,
  Mills, Marshall, Dwyer-Joyce, and Connolley]{mostafavi2017dynamic}
M.~Mostafavi, D.~Collins, M.~Peel, C.~Reinhard, S.~Barhli, R.~Mills,
  M.~Marshall, R.~Dwyer-Joyce, and T.~Connolley.
\newblock Dynamic contact strain measurement by time-resolved stroboscopic
  energy dispersive synchrotron x-ray diffraction.
\newblock \emph{Strain}, 53\penalty0 (2):\penalty0 e12221, 2017.

\bibitem[Ramadhan et~al.(2018)Ramadhan, Kockelmann, Tremsin, and
  Fitzpatrick]{ramadhan2018neutron}
R.~Ramadhan, W.~Kockelmann, A.~S. Tremsin, and M.~Fitzpatrick.
\newblock Neutron {T}ransmission {S}train measurements on {IMAT}: {R}esidual
  {S}train {M}apping in an {AlS}i{C}p {M}etal {M}atrix {C}omposite.
\newblock In \emph{MECA SENS 2017}. Materials Research Forum LLC, 2018.

\bibitem[Ramadhan et~al.(2019)Ramadhan, Kockelmann, Minniti, Chen, Parfitt,
  Fitzpatrick, and Tremsin]{ramadhan2019characterization}
R.~S. Ramadhan, W.~Kockelmann, T.~Minniti, B.~Chen, D.~Parfitt, M.~E.
  Fitzpatrick, and A.~S. Tremsin.
\newblock Characterization and application of {B}ragg-edge transmission imaging
  for strain measurement and crystallographic analysis on the {IMAT} beamline.
\newblock \emph{Journal of Applied Crystallography}, 52\penalty0 (2), 2019.

\bibitem[Rasmussen and Williams(2006)]{rasmussen2006gaussian}
C.~E. Rasmussen and C.~K. Williams.
\newblock \emph{Gaussian processes for machine learning}, volume~1.
\newblock MIT press Cambridge, 2006.

\bibitem[Santisteban et~al.(2001)Santisteban, Edwards, Steuwer, and
  Withers]{santisteban2001time}
J.~R. Santisteban, L.~Edwards, A.~Steuwer, and P.~J. Withers.
\newblock Time-of-flight neutron transmission diffraction.
\newblock \emph{Journal of applied crystallography}, 34\penalty0 (3):\penalty0
  289--297, 2001.

\bibitem[Santisteban et~al.(2002{\natexlab{a}})Santisteban, Edwards,
  Fitzpatrick, Steuwer, Withers, Daymond, Johnson, Rhodes, and
  Schooneveld]{santisteban02b}
J.~R. Santisteban, L.~Edwards, M.~E. Fitzpatrick, A.~Steuwer, P.~J. Withers,
  M.~R. Daymond, M.~W. Johnson, N.~Rhodes, and E.~M. Schooneveld.
\newblock Strain imaging by {B}ragg edge neutron transmission.
\newblock \emph{Nuclear Instruments and Methods in Physics Research Section A:
  Accelerators, Spectrometers, Detectors and Associated Equipment},
  481\penalty0 (1):\penalty0 765--768, 2002{\natexlab{a}}.

\bibitem[Santisteban et~al.(2002{\natexlab{b}})Santisteban, Edwards,
  Fizpatrick, Steuwer, and Withers]{santisteban02}
J.~R. Santisteban, L.~Edwards, M.~E. Fizpatrick, A.~Steuwer, and P.~J. Withers.
\newblock Engineering applications of {B}ragg-edge neutron transmission.
\newblock \emph{Applied Physics A}, 74\penalty0 (1):\penalty0 s1433--s1436,
  2002{\natexlab{b}}.

\bibitem[Sato et~al.(2011)Sato, Kamiyama, and Kiyanagi]{sato2011rietveld}
H.~Sato, T.~Kamiyama, and Y.~Kiyanagi.
\newblock A {R}ietveld-type analysis code for pulsed neutron bragg-edge
  transmission imaging and quantitative evaluation of texture and
  microstructure of a welded $\alpha$-iron plate.
\newblock \emph{Materials transactions}, pages 1105161372--1105161372, 2011.

\bibitem[Shinohara and Kai(2015)]{shinohara2015commissioning}
T.~Shinohara and T.~Kai.
\newblock Commissioning start of energy-resolved neutron imaging system,
  {RADEN} in {J}-{PARC}.
\newblock \emph{Neutron news}, 26\penalty0 (2):\penalty0 11--14, 2015.

\bibitem[Shinohara et~al.(2016)Shinohara, Kai, Oikawa, Segawa, Harada,
  Nakatani, Ooi, Aizawa, Sato, Kamiyama, et~al.]{shinohara2016final}
T.~Shinohara, T.~Kai, K.~Oikawa, M.~Segawa, M.~Harada, T.~Nakatani, M.~Ooi,
  K.~Aizawa, H.~Sato, T.~Kamiyama, et~al.
\newblock Final design of the energy-resolved neutron imaging system
  ``{RADEN}'' at {J}-{PARC}.
\newblock In \emph{Journal of Physics: Conference Series}, volume 746, page
  012007. IOP Publishing, 2016.

\bibitem[Tremsin et~al.(2011)Tremsin, McPhate, Kockelmann, Vallerga, Siegmund,
  and Feller]{tremsin11}
A.~S. Tremsin, J.~B. McPhate, W.~Kockelmann, J.~V. Vallerga, O.~H.~W. Siegmund,
  and W.~B. Feller.
\newblock High resolution {B}ragg edge transmission spectroscopy at pulsed
  neutron sources: {P}roof of principle experiments with a neutron counting
  {MCP} detector.
\newblock \emph{Nuclear Instruments and Methods in Physics Research Section A:
  Accelerators, Spectrometers, Detectors and Associated Equipment},
  633:\penalty0 S235--S238, 2011.

\bibitem[Tremsin et~al.(2012)Tremsin, McPhate, Steuwer, Kockelmann,
  M.~Paradowska, Kelleher, Vallerga, Siegmund, and Feller]{tremsin12}
A.~S. Tremsin, J.~B. McPhate, A.~Steuwer, W.~Kockelmann, A.~M.~Paradowska,
  J.~F. Kelleher, J.~V. Vallerga, O.~H.~W. Siegmund, and W.~B. Feller.
\newblock High-{R}esolution {S}train {M}apping {T}hrough {T}ime-of-{F}light
  {N}eutron {T}ransmission {D}iffraction with a {M}icrochannel {P}late
  {N}eutron {C}ounting {D}etector.
\newblock \emph{Strain}, 48\penalty0 (4):\penalty0 296--305, 2012.

\bibitem[Tremsin et~al.(2016)Tremsin, Gao, Dial, Grazzi, and
  Shinohara]{tremsin2016investigation}
A.~S. Tremsin, Y.~Gao, L.~C. Dial, F.~Grazzi, and T.~Shinohara.
\newblock Investigation of microstructure in additive manufactured inconel 625
  by spatially resolved neutron transmission spectroscopy.
\newblock \emph{Science and Technology of Advanced Materials}, 17\penalty0
  (1):\penalty0 324--336, 2016.
\newblock \doi{10.1080/14686996.2016.1190261}.

\bibitem[Vogel(2000)]{vogel2000tof}
S.~Vogel.
\newblock \emph{A {R}ietveld-{A}pproach for the {A}nalysis of {N}eutron
  {T}ime-{O}f-{F}light {T}ransmission {D}ata}.
\newblock PhD thesis, Christian-Albrechts-Universit{\"a}t, 2000.

\bibitem[Wahlstr{\"o}m(2015)]{wahlstrom2015modeling}
N.~Wahlstr{\"o}m.
\newblock \emph{Modeling of {M}agnetic {F}ields and {E}xtended {O}bjects for
  {L}ocalization {A}pplications}.
\newblock PhD thesis, Link{\"o}ping University Electronic Press, 2015.

\bibitem[Withers and Bhadeshia(2001)]{withers2001bresidual}
P.~J. Withers and H.~Bhadeshia.
\newblock Residual stress. {P}art 2--{N}ature and origins.
\newblock \emph{Materials science and technology}, 17\penalty0 (4):\penalty0
  366--375, 2001.

\bibitem[Wright and Nocedal(1999)]{wright1999numerical}
S.~Wright and J.~Nocedal.
\newblock Numerical optimization.
\newblock \emph{Springer Science}, 35\penalty0 (67-68):\penalty0 7, 1999.

\end{thebibliography}

\end{document}